\documentclass[aps,pra, twocolumn, amsmath, amssymb, showpacs, superscriptaddress]{revtex4}

\usepackage{graphicx}
\usepackage{bm}

\usepackage{xcolor}

\begin{document}

\title{An effective model for describing coherent population trapping resonances, which correctly takes into account the off-resonant frequency components in periodically modulated laser field}

\author{V.\,I.\,Yudin}
\email{viyudin@mail.ru}
\affiliation{Institute of Laser Physics SB RAS, pr. Akademika Lavrent'eva 15 B, Novosibirsk, 630090, Russia}
\affiliation{Novosibirsk State University, ul. Pirogova 1, Novosibirsk, 630090, Russia}
\affiliation{Novosibirsk State Technical University, pr. Karla Marksa 20, Novosibirsk, 630073, Russia}
\author{M.\,Yu.\,Basalaev}
\affiliation{Institute of Laser Physics SB RAS, pr. Akademika Lavrent'eva 15 B, Novosibirsk, 630090, Russia}
\affiliation{Novosibirsk State University, ul. Pirogova 1, Novosibirsk, 630090, Russia}
\affiliation{Novosibirsk State Technical University, pr. Karla Marksa 20, Novosibirsk, 630073, Russia}
\author{A.\,V.\,Taichenachev}
\affiliation{Institute of Laser Physics SB RAS, pr. Akademika Lavrent'eva 15 B, Novosibirsk, 630090, Russia}
\affiliation{Novosibirsk State University, ul. Pirogova 1, Novosibirsk, 630090, Russia}
\author{O.\,N.\,Prudnikov}
\affiliation{Institute of Laser Physics SB RAS, pr. Akademika Lavrent'eva 15 B, Novosibirsk, 630090, Russia}
\affiliation{Novosibirsk State University, ul. Pirogova 1, Novosibirsk, 630090, Russia}
%


\begin{abstract}
We have developed and tested an effective mathematical model for calculating the coherent population trapping (CPT) resonance in a periodically modulated laser field, when the modulation frequency $f$ varies near the fractional part of hyperfine splitting frequency in the ground state $\Delta_{\rm hfs}/N$ (where $N=1,2,3,...$). In such a polychromatic light, only two frequency components that are most resonant with the working optical transitions in the atom are accurately taken into account, while all other off-resonant frequency components are taken into account using the second-order perturbation theory in the field. Within the framework of the presented concept, equation for the atomic density matrix is obtained, in which taking into account of all off-resonant frequency components is reduced to the appearance of two new operators (non-diagonal, in the general case): the shift operator and the relaxation operator. In the case of three-level $\Lambda$-system, the adequacy of the presented effective model was verified by numerical calculations of various dependencies, in which we did not find visual differences from the exact calculations.

In addition to a significant mathematical simplification, our model provides a clear physical picture of various features of CPT spectroscopy in a periodically modulated laser field, including effects that have not been discussed in the scientific literature before. In particular, we show that the widespread viewpoint that the CPT resonance shift is determined by the usual ac Stark shifts of the lower levels is, in general, fundamentally incorrect, since the contribution to the light shift due to beats at the frequency $\Delta_{\rm hfs}$ between different off-resonant frequency components can be comparable to (or even dominate) with respect to the value of the standard ac Stark shift. Therefore, even if we have detailed information on the spectral composition of the modulated field (e.g. using a spectrum analyzer), this is, in general, absolutely insufficient to determine the light shift of the CPT resonance.
\end{abstract}

\maketitle

\section{Introduction}

Miniature atomic clocks (including chip-scale atomic clocks, CSACs), based on coherent population trapping (CPT), have been actively developed in many research centers around the world since the 2000s \cite{Kitching_IEEE_2000, Lutwak_PTTI_2001, Kitching_APL_2002, Stahler_OL_2002, Knappe_APL_2004,Vanier_APB_2005,Mescher_2005, Knappe_OptExp_2005,Shah_2010,Wang_ChinPhysB_2014, Kitching_ApplPhysRev_2018, Martinez_Nature_2023}. Currently, there are commercially available CSACs from various fabricators, which combine small dimensions and low power consumption with relatively high metrological characteristics  \cite{Cash_EFTF_2018,Marlow_IEEE_2021,MAC_site, NAC_site}. Such devices are widely used in various fields of science and technology \cite{Ramsey_2002, Maleki_Metrologia_2005, Riehle_FreqStandards_2005,Grewal_2007,Prestage_2007,Vanier_2015,Bock_2016,Kitching_ApplPhysRev_2018,Bandi_BEMSRep_2023}. At the same time, researches and development aimed at improving the metrological characteristics of CSACs continue  (e.g. see Refs.\,\cite{Hafiz_APL_2018, Skvortsov_QE_2020, Yun_PRAppl_2023}.

Initially, the CPT effect was discovered and theoretically developed in the case of a bichromatic field, when the frequency difference between its components varied around the hyperfine splitting frequency $\Delta_{\rm hfs}$ in the ground state of alkali metal atoms \cite{Alzetta_NCB_1976, Agapev_PhysUsp_1993, Arimondo_ProgOpt_1996}. However, modern miniature atomic clocks typically employ modulated field from a vertical-cavity semiconductor laser (VCSEL) to excite the CPT resonances in alkali metal vapors cells. Experimental and theoretical studies have shown that the polychromatic emission spectrum of a VCSEL, whose injection current is modulated at a frequency corresponding to the $\Delta_{\rm hfs}$ interval (usually in rubidium or cesium), is quite complex and depends on the individual design features of a particular laser. In addition, electro-optical and acousto-optical modulators are also often used for the modulation. It is obvious that a full theoretical model describing the formation of CPT resonances in such a polychromatic field should take into account, in general, the features of the modulated laser spectrum, namely, the amplitudes and phases of various frequency components. At the same time, in the existing literature, simplified concepts are typically used to describe the CPT resonances in a polychromatic laser field formed by a VCSEL with periodically modulated injection current. For example, a very simplified model has often been and continues to be used (e.g. see Refs.\,\cite{Vanier_APB_2005,Taichenachev_PRA_2003,Knappe_APB_2003,Yano_PRA_2014,Sokolov_JETP_2023}), which accurately takes into account only the two most resonance components, i.e. when the real polychromatic light is replaced by a two-frequency field, while all other spectral components are considered only in the context of the standard ac Stark shift of atomic levels. An ideal model of pure phase harmonic modulation, when the amplitudes and phases of the frequency components are completely determined by the corresponding Bessel functions, is also widely used. There are other options, such as mixed phase-amplitude harmonic modulation.

At the same time, one of the most important parameters that determines the metrological characteristics (long-term stability) of atomic CPT clocks is the light shift of the reference resonance. Therefore, the study of this shift and the development of methods for its suppression are constantly one of the most pressing problems of precision laser spectroscopy and quantum metrology \cite{Hafiz_APL_2018,Zhu_2000,Levi_IEEE_2000,Phillips_2005,Shah_2006,Yin_2017,Pollock_PRA_2018,Yudin_PRAppl_2020,Hafiz_PRAppl_2020,Chuchelov_2020,Pollock_APL_2022,Yudin_PRA_2023,Radnatarov_JETPLett_2023,Tsygankov_2024}. From physical considerations, it is obvious that the main parameters of the CPT resonance (amplitude and light shift of the resonance position), on which the metrological characteristics of miniature atomic clocks depend (stability and accuracy), are mainly determined by the spectral composition of the laser field. This fact is being used in experimental work to improve the characteristics of CPT resonances, including suppression of the light shift. In this case, greatly simplified ideas about the nature of this shift are used, intuitively believing that it is primarily determined by the ac Stark shift of the atomic levels in the ground state. However, as will be shown in our paper, such ideas, in general, are very far from reality and, therefore, a more accurate theoretical analysis is required.

In this paper, we develop and test an effective mathematical approach for calculating the CPT resonance in a periodically modulated laser field, when the modulation frequency $f$ varies near the fractional part of hyperfine splitting in the ground state $\Delta_{\rm hfs}/N$ (where $N = 1, 2, 3, ...$). In the presented model, two most resonant frequency components are taken into account exactly, while all other off-resonant frequency components are taken into account within the perturbation theory up to the second-order terms in the field in the equation for the atomic density matrix. Within the framework of this concept, we obtain an equation for the atomic density matrix, in which taking into account all off-resonant frequency components was reduced to the appearance of two new operators (non-diagonal, in the general case): the shift operator and the relaxation operator. In the case of a three-level $\Lambda$-system, the adequacy of the presented effective model is verified by comparison of graphs of various dependencies calculated on our method with exact calculations. In addition to a significant mathematical simplification (compared to exact calculation methods), our model gives a clear physical picture of various features of CPT spectroscopy in a periodically modulated laser field, including effects that have not been previously discussed in the scientific literature.

\section{General formalism}

\begin{figure}[t]
    \includegraphics[width=0.7\linewidth]{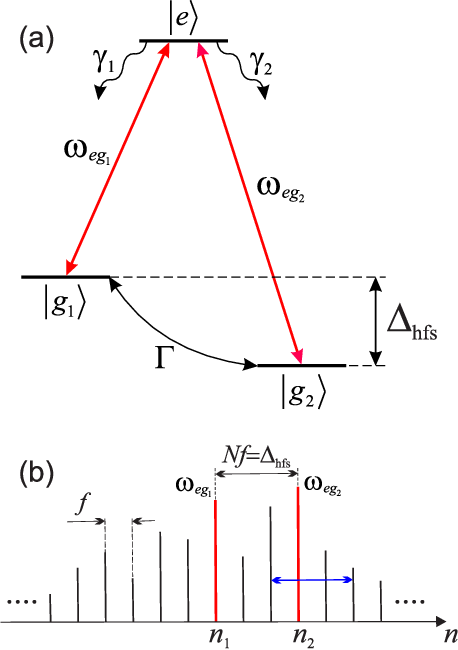}
    \caption{(a) Energy level diagram for the $\Lambda$ system. (b) Schematic representation of the spectral decomposition of the periodically modulated field (\ref{E_field}), where the red lines highlight the two frequency components resonant with the optical transitions in the $\Lambda$ system [see figure (a)].}\label{image:Lambda_scheme}
\end{figure}

Let us consider the interaction of an atomic three-level $\Lambda$-system [see Fig.\,\ref{image:Lambda_scheme}(a)] with a laser field with arbitrary amplitude-phase (amplitude-frequency) periodic modulation at the modulation frequency $f$
\begin{equation}\label{E_mod}
E(t)=e^{-i\omega t}{\cal E}(t)e^{-i\varphi(t)}+c.c.,
\end{equation}
where ${\cal E}(t+T)={\cal E}(t)$ and $\varphi(t+T)=\varphi(t)$ are arbitrary periodic functions with the same period $T=2\pi/f$ for amplitude and phase, respectively; $\omega$ is the carrier frequency. Using Fourier analysis, such a field can be described as polychromatic coherent field of the form
\begin{eqnarray}\label{E_field}
&&E(t)=\sum^{+\infty}_{n=-\infty}{\cal E}_{n}e^{-i(\omega_nt+\phi_n)} + c.c.=  \\
&&{\cal E}_{n_1}e^{-i(\omega_{n_1}t+\phi_{n_1})}+{\cal E}_{n_2}e^{-i(\omega_{n_2}t+\phi_{n_2})}+\nonumber \\
&&\sum\limits_{n\neq n_{1,2}}{\cal E}_{n}e^{-i(\omega_nt+\phi_n)}+c.c.,\nonumber
\end{eqnarray}
where ${\cal E}_{n}$ and $\phi_n$ are, respectively, the amplitude and phase of the $n$-th spectral component of electric field with the frequency
\begin{equation}\label{omega}
\omega_n=\omega+nf.
\end{equation}
 Thus, adjacent frequencies are separated from each other by the same value $f=\omega_{n+1}-\omega_{n}$ [see Fig.\,\ref{image:Lambda_scheme}(b)], which we will assume $f\approx\Delta_{\rm hfs}/N$, where $\Delta_{\rm hfs}=\omega_{eg_2}-\omega_{eg_1}$ is the transition frequency between the lower states $|g_1\rangle$ and $|g_2\rangle$, and $N$ is some fixed integer ($N=1,2,3,...$).

 In the expression (\ref{E_field}), we have identified two frequency components $\omega_{n_1}$ and $\omega_{n_2}$, which are resonant to the optical transitions in the $\Lambda$-system [see Fig.\,\ref{image:Lambda_scheme}(a) and the frequency components highlighted in red in Fig.\,\ref{image:Lambda_scheme}(b)]: $\omega_{n_1}\approx\omega_{eg_1}$ and $\omega_{n_2}\approx\omega_{eg_2}$. The difference between these frequencies automatically varies around the splitting frequency of the lower levels, $(n_{2}-n_{1})f\approx\Delta_{\rm hfs}$, i.e. $n_{2}=n_{1}+N$. Our aim is to develop such a mathematical model for calculating the spectroscopic CPT signal, in which two resonant frequency components $\omega_{n_1}$ and $\omega_{n_2}$ are taken into account exactly, while all other off-resonant frequency components with indices $n\neq n_{1,2}$ are taken into account by perturbation theory.

The theoretical description of the $\Lambda$-system will be carried using the formalism of the density matrix $\hat{\rho}$, which in the basis of atomic states $\{|e\rangle, |g_{1}\rangle, |g_{2}\rangle\}$ has the form
\begin{equation}\label{rho_matrix}
    \hat{\rho} =
        \begin{pmatrix}
        \rho_{ee} & \rho_{eg_{1}} & \rho_{eg_{2}}\\
        \rho_{g_{1}e} & \rho_{g_{1}g_{1}} & \rho_{g_{1}g_{2}}\\
        \rho_{g_{2}e} & \rho_{g_{2}g_{1}} & \rho_{g_{2}g_{2}}
        \end{pmatrix},
\end{equation}
and satisfies the equation
\begin{equation}\label{DensityMatrixEq}
    \frac{\partial}{\partial t}\hat{\rho} + \hat{\Gamma}\{\hat{\rho}\} = - i[\hat{H}^{(0)},\hat{\rho}] - i[\hat{V}(t),\hat{\rho}],
\end{equation}
where the symbol $[\hat{A},\hat{B}]=(\hat{A}\hat{B}-\hat{B}\hat{A})$ denotes the commutator of the operators $\hat{A}$ and $\hat{B}$. In the expression (\ref{DensityMatrixEq}), $\hat{H}^{(0)}$ is the Hamiltonian of the unperturbed atom, the linear operator functional $\hat{\Gamma}\{\hat{\rho}\}$ describes the relaxation processes, and $\hat{V}(t)$ describes the atom-field interaction.

Let us introduce the following projection operators:
\begin{eqnarray}\label{proek}
&&\hat{\Pi}_{e}=|e\rangle\langle e|,\; \hat{\Pi}_{eg_1}=|e\rangle\langle g_1|,\;\hat{\Pi}_{eg_2}=|e\rangle\langle g_2|, \\
&&\hat{\Pi}_{g_1}=|g_1\rangle\langle g_1|,\;\; \hat{\Pi}_{g_2}=|g_2\rangle\langle g_2|,\;\; \hat{\Pi}_{g}=\hat{\Pi}_{g_1}+\hat{\Pi}_{g_2}.\nonumber
\end{eqnarray}
Then the Hamiltonian of an unperturbed atom can be represented as
\begin{equation}\label{H0}
\hat{H}^{(0)}=-\omega_{eg_1}\hat{\Pi}_{g_1}-\omega_{eg_2}\hat{\Pi}_{g_2},
\end{equation}
while the operator $ \hat{\Gamma}\{\hat{\rho}\}$ in the relaxation constant model has the form
\begin{align}\label{Gamma}
& \hat{\Gamma}\{\hat{\rho}\}=\gamma_{\rm sp}\hat{\Pi}_{e}\hat{\rho}\hat{\Pi}_{e}-\gamma_1\hat{\Pi}^{\dagger}_{eg_1}\hat{\rho}\hat{\Pi}^{}_{eg_1}-\gamma_2\hat{\Pi}^{\dagger}_{eg_2}\hat{\rho}\hat{\Pi}^{}_{eg_2}+\\
 &\gamma_{\rm opt}\big(\hat{\Pi}_{e}\hat{\rho}\hat{\Pi}_{g}+\hat{\Pi}_{g}\hat{\rho}\hat{\Pi}_{e}\big)+\Gamma \big(\hat{\Pi}^{}_{g}\hat{\rho}\hat{\Pi}^{}_{g}-\frac{\hat{\Pi}^{}_{g}}{2}{\rm Tr}\{\hat{\Pi}^{}_{g}\hat{\rho}\hat{\Pi}^{}_{g}\}\big),\nonumber
\end{align}
where we use the set of relaxation constants: $\gamma_{\rm sp},\gamma_{1,2},\gamma_{\rm opt},\Gamma$. In this case, $\gamma_{\rm sp}$ is the rate of spontaneous decay of the upper level, and the constants $\gamma_{1}$ and $\gamma_{2}$ describe the partial rates of spontaneous arrival to the lower states $|g_{1}\rangle$ and $|g_{2}\rangle$, respectively [see Fig.\,\ref{image:Lambda_scheme}(a)], for which the condition of closed $\Lambda$-system is following
\begin{equation}\label{close}
\gamma_{1}+\gamma_{2}=\gamma_{\rm sp}.
\end{equation}
The constant $\gamma_{\rm opt}$ describes the rate of optical coherence decay both due to spontaneous emission and due to collisions with a buffer gas, i.e. the condition $\gamma_{\rm opt}\geq \gamma_{\rm sp}/2$ must be satisfied. The constant $\Gamma$ describes the rate of diffusion-transit depolarization of atoms in the ground state to the equilibrium isotropic distribution $\propto\hat{\Pi}_g$. In the case of CPT spectroscopy for alkali metal atoms (e.g., $^{87}$Rb or $^{133}$Cs) in gas cells with a buffer gas, the following hierarchy of quantities is usually satisfied
\begin{equation}\label{GGG}
\Gamma\ll \gamma_{\rm sp}\ll\gamma_{\rm opt}<f.
\end{equation}
Moreover, we assume that the value of $\gamma_{\rm opt}$ also exceeds the Doppler linewidth (i.e. $\gamma_{\rm opt} > k\bar{v}$, where $k$ is the light wavevector and $\bar{v}$ is the thermal velocity of atoms), which allows us to use the model of motionless atoms.

In the rotation wave approximation, the operator of the dipole interaction $\hat{V}(t)=-\hat{d}E(t)$ ($\hat{d}$ is the dipole moment operator of an atom) with the laser field (\ref{E_field}) has the form
\begin{equation}\label{V_matrix}
 \hat{V}(t)=-
 \begin{pmatrix}
        0 & V_{eg_{1}} & V_{eg_{2}}\\
        V^{*}_{eg_{1}} & 0 & 0\\
        V^{*}_{eg_{2}} & 0 & 0
        \end{pmatrix},
\end{equation}
in which the matrix elements are defined as
\begin{equation}\label{V_matrix_el}
    V^{}_{eg_{j}}(t) =\sum\limits_{n} \Omega_{n}^{(j)}e^{-i(\omega_nt+\phi_n)},\quad (j=1,2),
\end{equation}
where
\begin{equation}\label{Rabi_def}
   \Omega_{n}^{(j)} = d_{eg_{j}}{\cal E}^{}_{n}/\hbar,
\end{equation}
are the corresponding Rabi frequencies for the $n$-th spectral component, and $d_{eg_{j}} = \langle e|\hat{d}| g_{j}\rangle$ is the matrix element of the electric dipole moment operator for the optical transition $|g_{j}\rangle\leftrightarrow|e\rangle$ ($j= 1,2$).

Using the formulas (\ref{proek})-(\ref{V_matrix_el}), from the general equation for the density matrix (\ref{DensityMatrixEq}) we obtain the following system of equations for its elements
\begin{widetext}
\begin{eqnarray}\label{EqDensMatrE}
&&\frac{\partial}{\partial t}\rho_{g_{1}g_{1}} + \frac{\Gamma}{2} (\rho_{g_{1}g_{1}} -\rho_{g_{2}g_{2}})/2 - \gamma_{1}\rho_{ee} = i\sum\limits_{n}\left\{\Omega_{n}^{(1)\ast}\rho_{eg_{1}}e^{i(\omega_nt+\phi_n)} - \Omega_{n}^{(1)}\rho_{g_{1}e}e^{-i(\omega_nt+\phi_n)}\right\},\nonumber\\
    &&\frac{\partial}{\partial t}\rho_{g_{2}g_{2}} +\frac{\Gamma}{2} (\rho_{g_{2}g_{2}} - \rho_{g_{1}g_{1}})/2 - \gamma_{2}\rho_{ee} = i\sum\limits_{n}\left\{\Omega_{n}^{(2)\ast}\rho_{eg_{1}}e^{i(\omega_nt+\phi_n)} - \Omega_{n}^{(2)}\rho_{g_{1}e}e^{-i(\omega_nt+\phi_n)}\right\},\nonumber\\
    &&\frac{\partial}{\partial t}\rho_{ee} + \gamma_{\rm sp}\rho_{ee} = i\sum\limits_{n}\left\{\Omega_{n}^{(1)}\rho_{g_{1}e} + \Omega_{n}^{(2)}\rho_{g_{2}e}\right\}e^{-i(\omega_nt+\phi_n)}- i\sum\limits_{n}\left\{\Omega_{n}^{(1)\ast}\rho_{eg_{1}} + \Omega_{n}^{(2)\ast}\rho_{eg_{2}}\right\}e^{i(\omega_nt+\phi_n)},\nonumber\\
    &&\frac{\partial}{\partial t}\rho_{g_{1}g_{2}} + (\Gamma + i\Delta_{\rm hfs})\rho_{g_{1}g_{2}} = i\sum\limits_{n}\left\{\Omega_{n}^{(1)\ast}\rho_{eg_{2}}e^{i(\omega_nt+\phi_n)} -\Omega_{n}^{(2)}\rho_{g_{1}e}e^{-i(\omega_nt+\phi_n)}\right\},\\
    &&\frac{\partial}{\partial t}\rho_{eg_{1}} + (\gamma_{\rm opt} + i\omega_{eg_1})\rho_{eg_{1}} = i\sum\limits_{n}\left\{\Omega_{n}^{(1)}\rho_{g_{1}g_{1}} + \Omega_{n}^{(2)}\rho_{g_{2}g_{1}} - \Omega_{n}^{(1)}\rho_{ee}\right\}e^{-i(\omega_nt+\phi_n)},\nonumber\\
    &&\frac{\partial}{\partial t}\rho_{eg_{2}} + (\gamma_{\rm opt} + i\omega_{eg_2})\rho_{eg_{2}} =  i\sum_{n}\left\{\Omega_{n}^{(1)}\rho_{g_{1}g_{2}} + \Omega_{n}^{(2)}\rho_{g_{2}g_{2}} - \Omega_{n}^{(2)}\rho_{ee}\right\}e^{-i(\omega_nt+\phi_n)},\nonumber\\
    &&\rho_{g_{2}g_{1}} =\rho_{g_{1}g_{2}}^{*}, \quad \rho_{g_{1}e} =\rho_{eg_{1}}^{*}, \quad  \rho_{g_{2}e} =\rho_{eg_{2}}^{*}. \nonumber
\end{eqnarray}
\end{widetext}
In addition, taking into account the conservation of the total population, the equations (\ref{EqDensMatrE}) must be supplemented with a normalization condition
\begin{equation}\label{NormCond}
    {\rm Tr}\,\hat{\rho} = \rho_{g_{1}g_{1}} + \rho_{g_{2}g_{2}} + \rho_{ee} = 1,
\end{equation}
which is satisfied by the closedness of the $\Lambda$-system (\ref{close}).

\section{Reduced resonance model}

From a general mathematical viewpoint, the exact solution of the equation (\ref{DensityMatrixEq}) for the field (\ref{E_field}) requires the use of Fourier analysis, when the density matrix is represented as a decomposition
\begin{equation}\label{rho_F}
\hat{\rho}(t)=\sum^{+\infty}_{n=-\infty}\hat{\rho}^{(n)}e^{inft}\,.
\end{equation}
However, as noted above, our goal is to develop a mathematical model for calculating the spectroscopic CPT signal, in which only two resonant frequency components $\omega_{n_1}$ and $\omega_{n_2}$ [see (\ref{E_field}) and Fig.\,\ref{image:Lambda_scheme}(b)] are taken into account exactly, while all other off-resonant frequency components with indices $n\neq n_{1,2}$ are taken into account within the framework of the second-order perturbation theory in the electric field.

Such a reduction algorithm is presented in Appendix, where for the weak saturation regime of optical transitions in the $\Lambda$-system we obtained the following equation for the density matrix [see Eqs.\,(\ref{DensityMatrixEq-bichr})-(\ref{PopOp})]:
\begin{align}\label{DensityMatrix_red}
&\frac{\partial}{\partial t}\hat{\rho} + \hat{\Gamma}\{\hat{\rho}\} = - i[\hat{H}^{(0)},\hat{\rho}] + i[\hat{V}_{\rm res},\hat{\rho}]-\\
&i[\hat{S}_{\rm sh},\hat{\Pi}_g\hat{\rho}\hat{\Pi}_g]-\{\hat{P},\hat{\Pi}_g\hat{\rho}\hat{\Pi}_g\}+\hat{\Pi}_e{\rm Tr}\{\hat{P},\hat{\Pi}_g\hat{\rho}\hat{\Pi}_g\},\nonumber
\end{align}
where the symbol $\{\hat{A},\hat{B}\}=(\hat{A}\hat{B}+\hat{B}\hat{A})$ denotes the anticommutator of the operators $\hat{A}$ and $\hat{B}$. In Eq.\,(\ref{DensityMatrix_red}), the operator of interaction with the laser field contains only two frequency components
\begin{align}\label{V_res}
&\hat{V}_{\rm res}(t)=\\
&\begin{pmatrix}
        0 & \Omega_{n_1}^{(1)}e^{-i\omega_{n_1}t-i\phi_{n_1}} & \Omega_{n_2}^{(2)}e^{-i\omega_{n_2}t-i\phi_{n_2}}\\
        \Omega_{n_1}^{(1)*}e^{i\omega_{n_1}t+i\phi_{n_1}} & 0 & 0\\
        \Omega_{n_2}^{(2)*}e^{i\omega_{n_2}t+i\phi_{n_2}} & 0 & 0
        \end{pmatrix},\nonumber
\end{align}
which are most resonant with the optical transitions $|g_{1}\rangle\leftrightarrow|e\rangle$ and $|g_{2}\rangle\leftrightarrow|e\rangle$ [see Fig.\,\ref{image:Lambda_scheme}(b)].

As can be seen from Eq.\,(\ref{DensityMatrix_red}), taking into account the perturbation theory for all other off-resonant frequency components in (\ref{E_field}) with numbers $n\neq n_{1,2}$ has been reduced to the appearance of two new Hermitian operators $\hat{S}^{\dagger}_{\rm sh}=\hat{S}^{}_{\rm sh}$ and $\hat{P}^{\dagger}=\hat{P}$. The operator
\begin{equation}\label{S}
\hat{S}_{\rm sh}(t)=\begin{pmatrix}
        0 & 0 & 0\\
       0 & S^{}_{11} & S^{}_{12}e^{-iNft}\\
       0 & S^{*}_{12}e^{iNft} & S^{}_{22}
        \end{pmatrix},
\end{equation}
has the physical meaning of the Hamilton operator of the frequency shift due to off-resonant field components (\ref{E_field}). In this case, the diagonal element $S_{11}$ is the sum of the ac Stark shifts of the lower level $|g_{1}\rangle$ from all frequency components with numbers $n\neq n_1$, while the diagonal element $S_{22}$ is the sum of the quadratic ac Stark shifts of another lower level $|g_{2}\rangle$ from all frequency components with numbers $n\neq n_2$:
\begin{equation}\label{S1122}
S_{11}=\sum\limits_{n\neq n_{1}}\frac{\delta^{(1)}_{n}\big|\Omega_{n}^{(1)}\big|^{2}}{\gamma_{\rm opt}^{2} +\big|\delta^{(1)}_{n}\big|^{2}},\quad S_{22}=\sum\limits_{n\neq n_{2}}\frac{\delta^{(2)}_{n}\big|\Omega_{n}^{(2)}\big|^{2}}{\gamma_{\rm opt}^{2} +\big|\delta^{(2)}_{n}\big|^{2}},
\end{equation}
where $\delta^{(j)}_{n}$ is the one-photon detuning of $n$-th frequency component from the optical transition $|g_{j}\rangle\leftrightarrow|e\rangle$
\begin{equation}\label{delta}
\delta^{(j)}_{n}=\omega+n\frac{\Delta_{\rm hfs}}{N}-\omega^{}_{eg_j},\quad (j=1,2),
\end{equation}
in which we neglected the small variation of the modulation frequency $f$ with respect to the value $\Delta_{\rm hfs}/N$, i.e., in (\ref{delta}) we set $f=\Delta_{\rm hfs}/N$. In addition, the shift operator (\ref{S}) also contains non-diagonal elements oscillating at the frequency $(\omega^{}_{n_2}-\omega^{}_{n_1})=Nf$, for which the complex amplitude of these oscillations is defined as
\begin{equation}\label{S12}
S_{12}=\sum_{n\neq n_{1}}\frac{\delta^{(1)}_{n} \Omega^{(1)*}_{n}\Omega^{(2)}_{n+N}e^{i(\phi^{}_n-\phi^{}_{n+N})}}{\gamma_{\rm opt}^{2} + \big|\delta^{(1)}_{n}\big|^{2}}.
\end{equation}
This oscillating contribution to the shift operator $\hat{S}_{\rm sh}$ is due to the beats between the corresponding off-resonant components at the frequency difference $(\omega^{}_{n+N}-\omega^{}_{n})=Nf$ [see the connection with the blue arrow in Fig.\,\ref{image:Lambda_scheme}(b)], which coincides with the frequency difference between the two resonant components $\omega^{}_{n_2}$ and $\omega^{}_{n_2}$. As is clearly seen from (\ref{S12}), this matrix element depends on the phase difference between the different frequency components.

\begin{figure}[t]
    \includegraphics[width=0.9\linewidth]{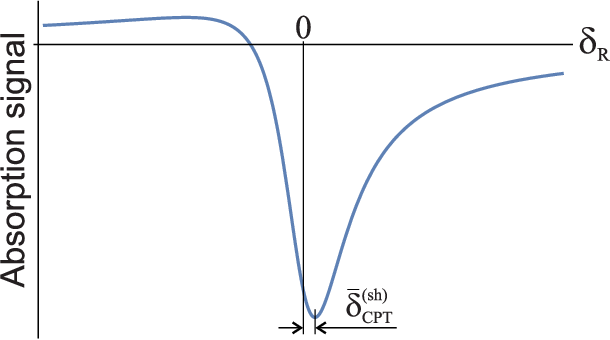}
    \caption{The typical lineshape of CPT resonance in the absorption signal of light passing through an atomic cell.}\label{lineshape}
\end{figure}

It should be especially noted that there is a very widespread viewpoint according to which the shift of resonance CPT is determined by the difference between usual ac Stark shifts of the lower levels $|g_{1}\rangle$ and $|g_{2}\rangle$:
\begin{eqnarray}\label{ac_Stark}
&&\bar{\delta}^{\rm (sh)}_\text{ac}=\sum\limits_{n}\left(\frac{\delta^{(1)}_{n}\big|\Omega_{n}^{(1)}\big|^{2}}{\gamma_{\rm opt}^{2} +\big|\delta^{(1)}_{n}\big|^{2}}-\frac{\delta^{(2)}_{n}\big|\Omega_{n}^{(2)}\big|^{2}}{\gamma_{\rm opt}^{2} +\big|\delta^{(2)}_{n}\big|^{2}}\right)=\nonumber\\
&&\frac{\delta^{(1)}_{n_1}\big|\Omega_{n_1}^{(1)}\big|^{2}}{\gamma_{\rm opt}^{2} +\big|\delta^{(1)}_{n_1}\big|^{2}}-\frac{\delta^{(2)}_{n_2}\big|\Omega_{n_2}^{(2)}\big|^{2}}{\gamma_{\rm opt}^{2} +\big|\delta^{(2)}_{n_2}\big|^{2}}+S_{11}-S_{22},
\end{eqnarray}
where the summation is carried out over all frequency components (including two resonant ones with indices $n= n_{1,2}$). However, as will be shown in the next sections, this viewpoint is fundamentally incorrect, since, in general, the contribution to the total shift of CPT resonance $\bar{\delta}^{\rm (sh)}_\text{CPT}$ (see Fig.\,\ref{lineshape}) due to the non-diagonal oscillating elements in (\ref{S}) can be comparable (or even dominate) to the value of $\bar{\delta}^{\rm (sh)}_\text{ac}$.

As for the other new operator $\hat{P}$ in Eq.\,(\ref{DensityMatrix_red}), which has the form
\begin{equation}\label{P}
\hat{P}=\begin{pmatrix}
        0 & 0 & 0\\
       0 & P^{}_{11} & P^{}_{12}e^{-iNft}\\
       0 & P^{*}_{12}e^{iNft} & P^{}_{22}
        \end{pmatrix},
\end{equation}
it is associated with the contribution of off-resonant frequency components to relaxation processes, which have some effect on the amplitude and shape of the CPT resonance. The expressions for real diagonal elements are written as
\begin{equation}\label{P1122}
P_{11}=\sum\limits_{n\neq n_{1}}\frac{\gamma_{\rm opt}\big|\Omega_{n}^{(1)}\big|^{2}}{\gamma_{\rm opt}^{2} +\big|\delta^{(1)}_{n}\big|^{2}},\quad P_{22}=\sum\limits_{n\neq n_{2}}\frac{\gamma_{\rm opt}\big|\Omega_{n}^{(2)}\big|^{2}}{\gamma_{\rm opt}^{2} +\big|\delta^{(2)}_{n}\big|^{2}},
\end{equation}
while the complex amplitude of the oscillating non-diagonal elements in (\ref{P}) is
\begin{equation}\label{P12}
P_{12}=\sum_{n\neq n_{1}}\frac{\gamma_{\rm opt}^{2} \Omega^{(1)*}_{n}\Omega^{(2)}_{n+N}e^{i(\phi^{}_n-\phi^{}_{n+N})}}{\gamma_{\rm opt}^{2} + \big|\delta^{(1)}_{n}\big|^{2}}.
\end{equation}
The physical meaning of the diagonal elements $P_{11}$ and $P_{22}$ consists of light-induced pumping from the lower states $|g_{1}\rangle$ and $|g_{2}\rangle$ to the upper state $|e\rangle$ due to the off-resonant spectral components, while the non-diagonal elements $P^{}_{12}=P^{*}_{21}$ additionally influence the formation of coherence in the ground state, described by the matrix elements $\rho_{g_1g_2}$ and $\rho_{g_2g_1}$.

Since the oscillations of the non-diagonal elements in the matrices $\hat{S}_{\rm sh}$ and $\hat{P}$ occur at the frequency that coincides with the difference between the resonant frequencies $(\omega_{n_2}-\omega_{n_1})=Nf$, then, having carried out the following replacement in the non-diagonal elements of the density matrix
\begin{eqnarray}\label{rho_res}
&&\rho_{eg_j}= e^{-i\omega_{n_j}t}\tilde{\rho}_{eg_j},\;\; \rho_{g_je}= e^{i\omega_{n_j}t}\tilde{\rho}_{g_je},\;\; (j=1,2),\nonumber\\
&&\rho_{g_1g_2}= e^{-iNft}\tilde{\rho}_{g_1g_2},\;\; \rho_{g_2g_1}= e^{iNft}\tilde{\rho}_{g_2g_1},
\end{eqnarray}
we finally obtain the equation for the modified density matrix $\hat{\tilde{\rho}}$:
\begin{align}\label{rho_res_eq}
&\frac{\partial}{\partial t}\hat{\tilde{\rho}} + \hat{\Gamma}\{\hat{\tilde{\rho}}\} = - i[\hat{H}^{(0)}_{\rm res},\hat{\tilde{\rho}}] + i[\hat{\widetilde{V}}_{\rm res},\hat{\tilde{\rho}}]-\\
&i[\hat{\widetilde{S}}_{\rm sh},\hat{\Pi}_g\hat{\tilde{\rho}}\hat{\Pi}_g]- \{\hat{\widetilde{P}},\hat{\Pi}_g\hat{\tilde{\rho}}\hat{\Pi}_g\}+\hat{\Pi}_e{\rm Tr}\{\hat{\widetilde{P}},\hat{\Pi}_g\hat{\tilde{\rho}}\hat{\Pi}_g\},\nonumber\\
&  {\rm Tr}\{\hat{\tilde{\rho}}\} = \rho_{g_{1}g_{1}} + \rho_{g_{2}g_{2}} + \rho_{ee} = 1,\nonumber
\end{align}
in which there are no oscillating terms. In this case, the following expressions for the operators in (\ref{rho_res_eq}) take place:
\begin{eqnarray}\label{H0_res}
&&\hat{H}^{(0)}_{\rm res}=\delta_1\hat{\Pi}_{g_1}+\delta_2\hat{\Pi}_{g_2}=\begin{pmatrix}
        0 & 0 & 0\\
       0 & \delta_1 & 0\\
       0 & 0 & \delta_2
        \end{pmatrix}, \\
&&\delta^{}_j=\omega^{}_{n_j}-\omega^{}_{eg_j}=\omega+fn_j-\omega^{}_{eg_j},\quad (j=1,2),\nonumber
\end{eqnarray}
\begin{equation}\label{V_res2}
\hat{\widetilde{V}}_{\rm res}=
\begin{pmatrix}
        0 & \Omega_{n_1}^{(1)}e^{-i\phi_{n_1}} & \Omega_{n_2}^{(2)}e^{-i\phi_{n_2}}\\
        \Omega_{n_1}^{(1)*}e^{i\phi_{n_1}} & 0 & 0\\
        \Omega_{n_2}^{(2)*}e^{i\phi_{n_2}} & 0 & 0
        \end{pmatrix},
\end{equation}
\begin{equation}\label{S_res}
\hat{\widetilde{S}}_{\rm sh}=\begin{pmatrix}
        0 & 0 & 0\\
       0 & S^{}_{11} & S^{}_{12}\\
       0 & S^{*}_{12} & S^{}_{22}
        \end{pmatrix},
\end{equation}
\begin{equation}\label{P_res}
\hat{\widetilde{P}}=\begin{pmatrix}
        0 & 0 & 0\\
       0 & P^{}_{11} & P^{}_{12}\\
       0 & P^{*}_{12} & P^{}_{22}
        \end{pmatrix}.
\end{equation}
In the calculations, we will use the general one-photon detuning for both resonant components ($n=n_{1,2}$), which we define as follows
\begin{equation}\label{1-ph_def}
    \delta_\text{1-ph} = \delta^{}_{1}\big|_{f=\Delta_{\rm hfs}/N}= \delta^{}_{2}\big|_{f=\Delta_{\rm hfs}/N}=\omega+n_1\frac{\Delta_{\rm hfs}}{N}-\omega_{eg_1}.
\end{equation}
As a variable two-photon (Raman) detuning we will use the value
\begin{equation}\label{2-ph_def}
 \delta_{\rm R} = \omega_{n_2}-\omega_{n_1}-\Delta_{\rm hfs}=Nf-\Delta_{\rm hfs}\,.
\end{equation}
In this case, according to the definitions (\ref{1-ph_def})-(\ref{2-ph_def}), the detunings $\delta_1$ and $\delta_2$ in (\ref{H0_res}) are rewritten as
\begin{equation}\label{d1d2}
  \delta_1=\delta_\text{1-ph}+\frac{n_1}{N} \delta_{\rm R},\quad \delta_2=\delta_\text{1-ph}+\frac{n_2}{N} \delta_{\rm R}\,,
\end{equation}
where $n_2-n_1=N$. As a spectroscopic signal describing the CPT resonance, we will consider the functional dependence of the upper level population $|e\rangle$ on the two-photon detuning, $\rho_{ee}(\delta_{\rm R})$.
This dependence corresponds to the CPT resonance in the absorption signal of light passing through an atomic cell, the typical lineshape of which is shown in Fig.\,\ref{lineshape}. Below we will perform a numerical test of the solutions of Eq.\,(\ref{rho_res_eq}) for different regimes.

\section{Numerical verification of reduced model}

Let us perform a comparative analysis of the steady state solution of Eq.\,(\ref{rho_res_eq}) [i.e., setting $d/dt\,\hat{\tilde{\rho}}=0$ in (\ref{rho_res_eq})] with the exact solution of the basic equation (\ref{DensityMatrixEq}), which is found numerically using the Fourier expansion of the initial density matrix $\hat{\rho}$ in harmonics $e^{inft}$ [see (\ref{rho_F})]. In the last case, as a spectroscopic signal we consider the functional dependence of the upper level population on the two-photon detuning for the zero harmonic, $\rho^{(0)}_{ee}(\delta_{\rm R})$.

\begin{figure}[t]
    \includegraphics[width=0.95\linewidth]{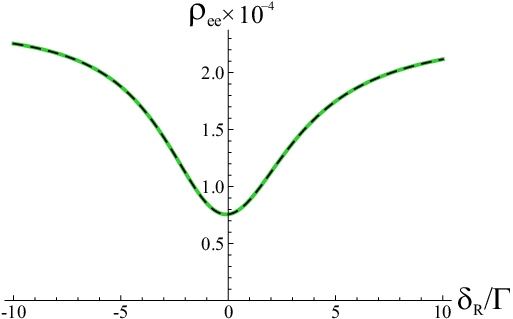}
    \caption{\label{image:PhM-AM} Calculation of CPT resonance lineshape in the case of the harmonic amplitude-phase modulation (\ref{E_am-ph}). The green solid and black dashed lines (which visually coincide) correspond to calculation based on the effective model (\ref{rho_res_eq}) and the exact calculation for the full equation (\ref{DensityMatrixEq}). Numerical parameters of the calculation:  $\bar{\Omega}^{(1)}/\gamma_{\rm sp} =\bar{\Omega}^{(2)}/\gamma_{\rm sp} = 0.25$, $\gamma_{\rm opt}/\gamma_{\rm sp} = 50$, $\gamma_1/\gamma_{\rm sp}=\gamma_2/\gamma_{\rm sp}=1/2$, $\Gamma/\gamma_{\rm sp} = 10^{-4}$, $\Delta_{\rm hfs}/\gamma_{\rm sp} = 1188.64$, $\delta_\text{1-ph}/\gamma_{\rm sp} = 25$, $M=3$, $A = 0.5$, $\theta = \pi/3$.}
\end{figure}

To do this comparison, we consider a general case of harmonic amplitude-phase modulation, when the electric field has the form
\begin{equation}\label{E_am-ph}
    E = E_{0}[1 + A \cos{(f t +\theta)}]e^{-i(\omega t + M\sin{f t})} + \rm{c.c.},
\end{equation}
where $E_{0}$ is the laser field amplitude, $A$ is the amplitude modulation depth, $M$ is the phase modulation index, $\theta$ is the relative phase between the amplitude and phase modulations. To proceed to the description of CPT resonances using the effective model (\ref{rho_res_eq}), we expand the expression (\ref{E_am-ph}) into harmonics
\begin{eqnarray}\label{E_am-ph_harmonics}
    &&E = E_{0}\sum_{n = -\infty}^{\infty}\big(J_{n}(M) + \\
    &&\frac{A}{2}[J_{n - 1}(M)e^{-i\theta} + J_{n + 1}(M)e^{i\theta}]\big)e^{-i\omega_n t} + \rm{c.c.},\nonumber
\end{eqnarray}
where $J_{\beta}(M)$ is the Bessel function of the first kind of order $\beta$. In this case using (\ref{E_field}), (\ref{V_matrix_el}) and (\ref{Rabi_def}), we find the following expression for the Rabi frequency (together with the phase factor) of the $n$-th spectral component
\begin{eqnarray}\label{Rabi_am-ph}
    &&\Omega^{(j)}_{n}e^{-i\phi_n} =\bar{\Omega}^{(j)}\times\\
    &&\big(J_{n}(M) + \frac{A}{2}[J_{n - 1}(M)e^{-i\theta} + J_{n + 1}(M)e^{i\theta}]\big),\; (j=1,2)\nonumber,
\end{eqnarray}
which we will use to calculate the operators $\hat{\widetilde{S}}_{\rm sh}$ and $\hat{\widetilde{P}}$ [see formulas (\ref{S_res}) and (\ref{P_res})], where $\bar{\Omega}^{(j)}=d_{eg_{j}}E_{0}/\hbar$ ($j=1,2$). Usually in practice, either the field components with $n = -1,+1$ for modulation at frequency $f \approx \Delta_{\rm hfs}/2$ (i.e. $N=2$) or the field components with $n = 0,+1$ (or $n = -1,0$) for modulation at frequency $f \approx \Delta_{\rm hfs}$ (i.e. $N=1$) are in resonance with optical transitions $|e\rangle \leftrightarrow |g_{1}\rangle$ and $|e\rangle \leftrightarrow |g_{2}\rangle$. In the calculations, we assume that the dipole moments of optical transitions are the same, $d_{eg_{1}} = d_{eg_{2}} = d$, i.e. $\bar{\Omega}^{(1)} =\bar{\Omega}^{(2)}$. This occurs, for example, in the CPT spectroscopy  with a circularly polarized field at D1 line of alkali metal atoms (Rb, Cs), when Zeeman sublevels with magnetic quantum number $m=0$ at the lower hyperfine levels are used.

\begin{figure}[t]
\includegraphics[width=0.8\linewidth]{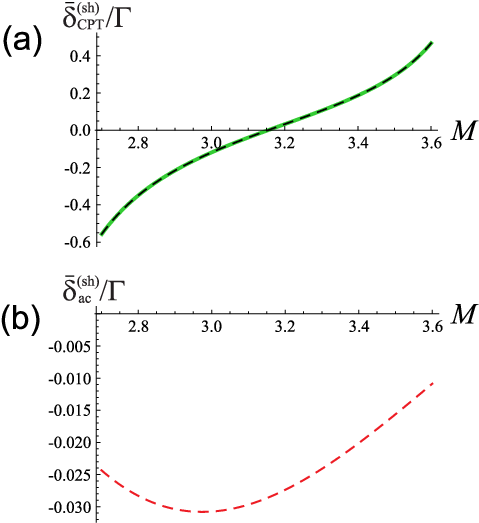}
\caption{\label{image:Shift_vs_M} Dependence of CPT resonance peak shift $\bar{\delta}^{\rm (sh)}_\text{CPT}$ versus the phase modulation index $M$: (a) The green solid and black dashed lines (which visually coincide) correspond to calculation based on the effective model (\ref{rho_res_eq}) and the exact calculation for the full equation (\ref{DensityMatrixEq}). (b) Calculation of the standard ac Stark shift $\bar{\delta}^{\rm (sh)}_\text{ac}$ of the reference rf transition [see Eq.\,(\ref{ac_Stark})]. Numerical parameters of the calculation: $\bar{\Omega}^{(1)}/\gamma_{\rm sp} =\bar{\Omega}^{(2)}/\gamma_{\rm sp} = 0.25$, $\gamma_{\rm opt}/\gamma_{\rm sp} = 50$, $\gamma_1/\gamma_{\rm sp}=\gamma_2/\gamma_{\rm sp}=1/2$, $\Gamma/\gamma_{\rm sp} = 10^{-4}$, $\Delta_{\rm hfs}/\gamma_{\rm sp} = 1188.64$, $\delta_\text{1-ph} = 0$, $A = 0$.}
\end{figure}

Let us first consider the case of the modulation frequency $f \approx\Delta_{\rm hfs}$ (i.e. $N=1$), when the carrier frequency $\omega=\omega_0$ is resonant with the optical transition $|e\rangle \leftrightarrow |g_{1}\rangle$ ($\omega=\omega_{eg_1}$), and the first harmonic $\omega_{+1}=\omega+f$ varies near the frequency $\omega_{eg_2}$ of the optical transition $|e\rangle \leftrightarrow |g_{2}\rangle$. The Fig.\,\ref{image:PhM-AM} shows a comparison of the CPT resonance lineshape calculated based on: (A) the exact solution of Eq.\,(\ref{DensityMatrixEq}) for the field (\ref{E_am-ph}) (black dashed line); (B) the effective bichromatic model (\ref{rho_res_eq}) for the expansion of the field (\ref{E_am-ph_harmonics}) (green solid line). As can be seen, the calculation results for the effective model are in very good agreement with the exact solution, which confirms the adequacy of our approach.

However, from the metrological viewpoint, the most important characteristic of the CPT resonance is the light shift. Let us consider this problem in more detail. For example, in the case of the one-photon resonance (i.e. $\delta_\text{1-ph} = 0$) and pure phase modulation [i.e. $A=0$ in (\ref{E_am-ph})], from formulas (\ref{S1122})-(\ref{S12}) and (\ref{Rabi_am-ph}) one can obtain simple analytical expressions for the matrix elements of the shift operator $\hat{\widetilde{S}}_{\rm sh}$
\begin{eqnarray}\label{ShiftOp_Full}
    &&S_{11} = 0, \quad  S_{22} = -\frac{2|\bar{\Omega}^{(2)}|^{2} J_{0}(M)J_{+1}(M)}{M\Delta_{\rm hfs}}, \nonumber\\
    &&S^{}_{12} = S^{*}_{21} = \frac{\bar{\Omega}^{(1)}\bar{\Omega}^{(2)*}[1-J_{0}^{2}(M)]}{M\Delta_{\rm hfs}},
\end{eqnarray}
From Eq.\,(\ref{ShiftOp_Full}), we can see that it is impossible to select the phase modulation index $M$ for which the standard ac Stark shift $\bar{\delta}^{\rm (sh)}_\text{ac}$ [see Eq.\,(\ref{ac_Stark})] of the rf transition $|g_{1}\rangle \leftrightarrow |g_{2}\rangle$ is absent and simultaneously the amplitudes of both resonant harmonics are non-zero, ${\cal E}_{0,+1}\neq 0$. Nevertheless, calculations show that there is such an index $M$ for which the resonance top does not experience the light shift for non-zero standard ac Stark shift $\bar{\delta}^{\rm (sh)}_\text{ac}$. Indeed, this is clearly seen from Fig.\,\ref{image:Shift_vs_M}, which contains the dependence of the resonance peak shift on the phase modulation index $M$ for the effective model [see Fig.\,\ref{image:Shift_vs_M}(a)] and the dependence of the standard ac Stark shift $\bar{\delta}^{\rm (sh)}_\text{ac}$ [see Fig.\,\ref{image:Shift_vs_M}(b)]. The graphs show very strong difference both in the character of the dependence and in the shift value itself. Thus, in this spectroscopic scheme, the contribution from the non-diagonal elements $S^{}_{12}=S^{*}_{21}\neq 0$ of the shift operator $\hat{\widetilde{S}}_{\rm sh}$ is dominant.

\begin{figure}[t]
    \includegraphics[width=0.8\linewidth]{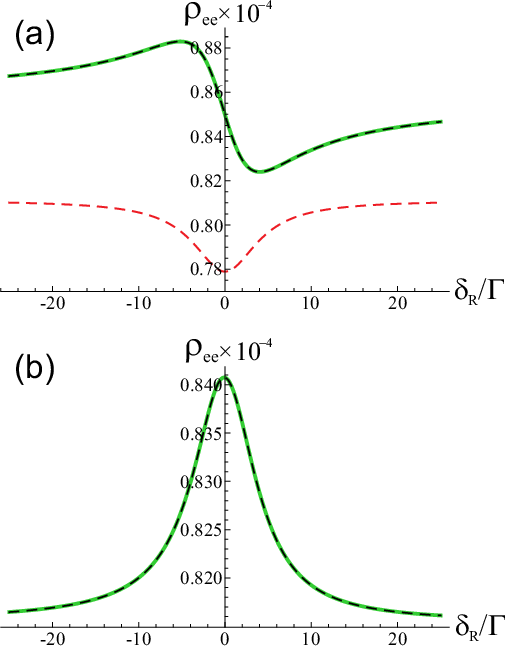}
    \caption{\label{image:PhM_LineShape} The CPT resonance lineshape: (a) phase modulation index $M = 2.35$; (b) phase modulation index $M = 2.405$. The green solid and black dashed lines (which visually coincide) correspond to calculation based on the effective model (\ref{rho_res_eq}) and the exact calculation for the full equation (\ref{DensityMatrixEq}), the red dashed line in figure (a) [absent in figure (b)] corresponds to the widely used approach, in which off-resonant spectral components are taken into account only as ac Stark shift of the reference rf transition. Numerical parameters of the calculation: $\bar{\Omega}^{(1)}/\gamma_{\rm sp} =\bar{\Omega}^{(2)}/\gamma_{\rm sp} = 0.25$, $\gamma_{\rm opt}/\gamma_{\rm sp} = 50$, $\gamma_1/\gamma_{\rm sp}=\gamma_2/\gamma_{\rm sp}=1/2$, $\Gamma/\gamma_{\rm sp} = 10^{-4}$, $\Delta_{\rm hfs}/\gamma_{\rm sp} = 1188.64$, $\delta_\text{1-ph} = 0$, $A = 0$.}
\end{figure}

However, the non-diagonal elements $S^{}_{12}=S^{*}_{21}\neq 0$ of the shift operator $\hat{\widetilde{S}}_{\rm sh}$ can have a significant effect not only on the resonance position, but also on the lineshape. Let us demonstrate this using an example where the modulation index is chosen so that the amplitude of one of the resonant components is small. The Fig.\,\ref{image:PhM_LineShape}(a) shows a radical change in the lineshape at $M = 2.35$, which cannot be described within the widespread approach taking into account only ac Stark shift [see the blue line in Fig.\,\ref{image:PhM_LineShape}(a)]. Moreover, our effective bichromatic model gives the correct result even in the case where the amplitude of one of the resonant components is zero, which is achieved for $M=2.405$. In this case, as shown in Fig.\,\ref{image:PhM_LineShape}(b), a change in the sign of CPT resonance occurs, while the CPT signal formed by the resonant components is completely absent [see the absence of the blue line in Fig.\,\ref{image:PhM_LineShape}(b)].

\begin{figure}[t]
    \includegraphics[width=0.8\linewidth]{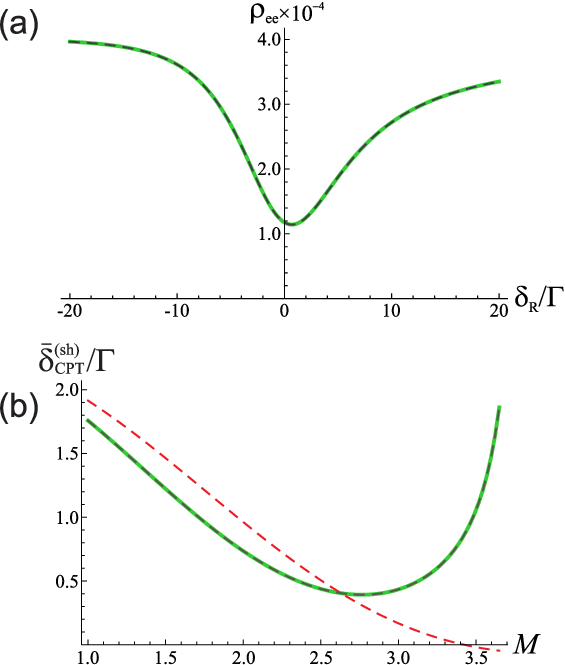}
    \caption{CPT resonance in the case of harmonic amplitude-phase modulation (\ref{E_am-ph}) at the modulation frequency $f\approx\Delta_{\rm hfs}/3$ (i.e. $N=3$), when the resonant harmonics are with indexes $n_1=-1$ and $n_2=+2$: (a) CPT resonance lineshape for the modulation parameters $M=3$, $A = 0.5$, $\theta = \pi/3$ and one-photon detuning $\delta_\text{1-ph} = 25$, where the green solid and black dashed lines (which visually coincide) correspond to calculation based on the effective model (\ref{rho_res_eq}) and the exact calculation for the full equation (\ref{DensityMatrixEq}); (b) the dependence of the resonance peak shift $\bar{\delta}^{\rm (sh)}_\text{CPT}$ versus the phase modulation index $M$ for $A=0$ and $\delta_\text{1-ph} = 0$, where the green solid and black dashed lines (which visually coincide) correspond to calculation based on the effective model (\ref{rho_res_eq}) and the exact calculation for the full equation (\ref{DensityMatrixEq}), and the red dashed line corresponds to the standard ac Stark shift $\bar{\delta}^{\rm (sh)}_\text{ac}$ [see Eq.\,(\ref{ac_Stark})]. Other numerical parameters of the calculation:
    $\bar{\Omega}^{(1)}/\gamma_{\rm sp} =\bar{\Omega}^{(2)}/\gamma_{\rm sp} = 0.25$, $\gamma_{\rm opt}/\gamma_{\rm sp} = 50$, $\gamma_1/\gamma_{\rm sp}=\gamma_2/\gamma_{\rm sp}=1/2$, $\Gamma/\gamma_{\rm sp} = 10^{-4}$, $\Delta_{\rm hfs}/\gamma_{\rm sp} = 1188.64$.}\label{image:Hm1p2}
\end{figure}

For the completeness, Fig.\,\ref{image:Hm1p2} shows calculations for the modulation frequency $f \approx\Delta_{\rm hfs}/3$ (i.e. $N=3$), when the resonant harmonics are $n_1=-1$ and $n_2=+2$. As can be clearly seen, the results of the calculation based on the effective model (\ref{rho_res_eq}) do not visually differ from the exact calculation for the full equation (\ref{DensityMatrixEq}). It is also obvious that the shift of the CPT resonance [see in Fig.\,\ref{image:Hm1p2}(b)] is poorly described by the standard ac Stark shift $\bar{\delta}^{\rm (sh)}_\text{ac}$, that is a consequence of the influence of the non-diagonal elements $S^{}_{12}=S^{*}_{21}\neq 0$ in the shift operator $\hat{\widetilde{S}}_{\rm sh}$.

\section{Modulation at frequency $\Delta_{\rm hfs}/2$}

In this section, we separately investigate the case of periodic modulation at the frequency $f\approx\Delta_{\rm hfs}/2$ (i.e. $N=2$). In this case, we assume that the carrier frequency $\omega=(\omega_{eg_1}+\omega_{eg_2})/2$ is tuned to the middle between the transitions $|e\rangle \leftrightarrow |g_{1}\rangle$ and $|e\rangle \leftrightarrow |g_{2}\rangle$, i.e. the spectral components with $n=\pm 1$ are resonant with the optical transitions in the $\Lambda$ system. In addition, we also assume $d_{eg_1}=d_{eg_2}=d$ and $\gamma_1=\gamma_2=\gamma_{\rm sp}/2$.
\begin{figure}[t]
    \includegraphics[width=0.95\linewidth]{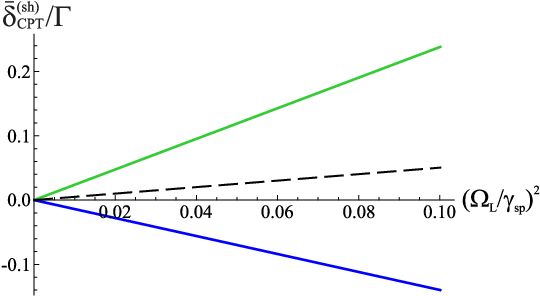}
    \caption{\label{image:Shift_vs_R0} Dependences of the light shift of CPT resonance versus the laser field intensity ($\propto\Omega^{2}_{L}$), calculated on the base of the effective model (\ref{rho_res_eq}), which completely coincide with the exact calculations for the full equation (\ref{DensityMatrixEq}). The blue solid line corresponds to the phases (\ref{phi_1}), the green solid line corresponds to the phases (\ref{phi_2}), and the red dashed line corresponds to the dependence of the standard ac Stark shift $\bar{\delta}^{\rm (sh)}_\text{ac}$ [see Eq.\,(\ref{ac_Stark})]. Numerical parameters of the calculation: $\gamma_{\rm opt}/\gamma_{\rm sp} = 50$, $\gamma_1/\gamma_{\rm sp}=\gamma_2/\gamma_{\rm sp}=1/2$, $\Gamma/\gamma_{\rm sp} = 10^{-4}$, $\Delta_{\rm hfs}/\gamma_{\rm sp} = 1188.64$, $\delta_\text{1-ph} = 0$.}
\end{figure}

Let us consider two variants of the field, with the same spectral composition of eleven frequency components ($-5\leq n\leq +5$) with a fixed ratio between the real amplitudes, which are chosen quite arbitrarily:
\begin{eqnarray}\label{E_example}
&& {\cal E}_{-5} = -0.1E_{0},\;{\cal E}_{-4} = 0.2E_{0},\;{\cal E}_{-3} = -0.6E_{0},\nonumber\\
&&{\cal E}_{-2} = 0.25E_{0},\;{\cal E}_{-1} = -0.7E_{0},\;{\cal E}_{0} = 0.15E_{0},\\
&&{\cal E}_{+1} = 0.4E_{0},\;{\cal E}_{+2} = 0.5E_{0},\;{\cal E}_{+3} = 0.35E_{0},\nonumber\\
&&{\cal E}_{+4} = 0.1E_{0},\;{\cal E}_{+5}= 0.25E_{0},\nonumber
\end{eqnarray}
where the parameter $E_{0}$ depends on the total intensity of the laser field $I\propto |E_{0}|^2$. In the first variant, the phases of all spectral components are equal to zero
\begin{equation}\label{phi_1}
 \phi_{n}=0,\quad (n=-5,-4,...,+4,+5) .
\end{equation}
In the second variant of the light field, the phases of the spectral components are different from each other and are determined as follows:
\begin{eqnarray}\label{phi_2}
&&\phi_{-5} = 5\pi/6,\;\;\phi_{-4} =  4\pi/6,\;\;\phi_{-3} =  3\pi/6,\;\;\phi_{-2} =  2\pi/6, \nonumber\\
&&\phi_{-1} = \pi/6,\;\;\phi_{0} = 0,\;\;\phi_{+1} =  -\pi/6,\;\;\phi_{+2} =  -2\pi/6,\nonumber\\
&&\phi_{+3} =  -3\pi/6,\;\; \phi_{+4} =  -4\pi/6,\;\;\phi_{+5} =  -5\pi/6,
\end{eqnarray}
where the phase difference between adjacent harmonics is $(\phi_{n+1}-\phi_{n})=-\pi/6$.

For demonstration, Fig.\,\ref{image:Shift_vs_R0} shows the dependences of the CPT resonance peak shift on the laser field intensity $I\propto |\Omega_{L}|^2$ (where $\Omega_{L}=dE_{0}/\hbar$) for two spectral compositions that differ only in phase relationships [see (\ref{phi_1}) and (\ref{phi_2})]. As it is seen, the field shifts differ radically not only in magnitude but also in sign. At the same time, the standard ac Stark shift $\bar{\delta}^{\rm (sh)}_\text{ac}$ is the same in both cases (see the black dashed line in Fig.\,\ref{image:Shift_vs_R0}).

Thus, based on the results of this and the previous (see Section~IV) sections, we have clearly shown that the widespread viewpoint according to which the CPT resonance shift is determined by the usual ac Stark shifts of the lower levels is fundamentally incorrect. Indeed, as was shown above, even if we have detailed information on the amplitudes of the spectral components of the modulated field (e.g. using a spectrum analyzer), this is, in general, absolutely insufficient for determining the light shift of the CPT resonance if there is no information on the phase relationship between the various components of the field.

\section{The case of equality of resonant components, $|\Omega_{n_{1}}| = |\Omega_{n_{2}}|$}

In addition, as a result of a detailed analysis, we have found an exceptional feature of the modulated field when the equality of the Rabi frequencies for two resonant components is fulfilled, $|\Omega_{n_{1}}| = |\Omega_{n_{2}}|$, the value of which should not be too small. In this case, the influence of the non-diagonal elements of the shift operator $S^{}_{12}=S^{*}_{21}\neq 0$ becomes insignificant and the light shift of the CPT resonance is determined primarily by the standard ac Stark shift $\bar{\delta}^{\rm (sh)}_\text{ac}$ of the reference rf transition $|g_{1}\rangle \leftrightarrow |g_{2}\rangle$. Mathematically, this result is due to the fact that in the equations for the rf coherence $\rho^{}_{g_{1}g_{2}}$ and $\rho^{*}_{g_{2}g_{1}}$ the non-diagonal elements of the shift operator $S^{}_{12}=S^{*}_{21}$ are multiplied by the population difference for the lower levels $(\rho_{g_{1}g_{1}}-\rho_{g_{2}g_{2}})$ [see in (\ref{EqDensMatrEls_F-nr})], which becomes small at $|\Omega_{n_{1}}| = |\Omega_{n_{2}}|$.

\begin{figure}[t]
    \includegraphics[width=0.95\linewidth]{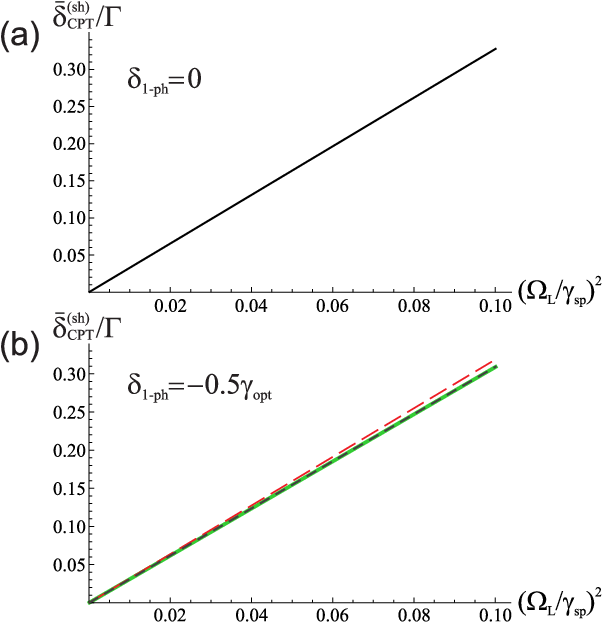}
    \caption{\label{image:Shift_vs_RR_equal} Dependences of the light shift of CPT resonance versus the laser intensity ($\sim\Omega_{L}^{2}$) for the field (\ref{E_example_2}) with phase relations (\ref{phi_1}) and (\ref{phi_2}) for two values of the one-photon detuning: (a)~$\delta_{\text{1-ph}} = 0$, where both graphs visually coincide with the dependence of the standard ac Stark shift $\bar{\delta}^{\rm (sh)}_\text{ac}$ [see Eq.\,(\ref{ac_Stark})]; (b)~$\delta_{\text{1-ph}} = -0.5\gamma_{\rm opt}$, where the black dashed and green solid lines (which visually coincide) are calculated for phase relations (\ref{phi_1}) and (\ref{phi_2}) using the effective model (\ref{rho_res_eq}), and the red dashed line corresponds to the standard ac Stark shift $\bar{\delta}^{\rm (sh)}_\text{ac}$ [see Eq.\,(\ref{ac_Stark})]. Numerical parameters of the calculations: $\gamma_{\rm opt}/\gamma_{\rm sp} = 50$, $\gamma_1/\gamma_{\rm sp}=\gamma_2/\gamma_{\rm sp}=1/2$, $\Gamma/\gamma_{\rm sp} = 10^{-4}$, $\Delta_{\rm hfs}/\gamma_{\rm sp} = 1188.64$.}
\end{figure}

To demonstrate this fact, consider a field modulated at the frequency $f\approx\Delta_{\rm hfs}/2$, for which the real amplitudes of the eleven spectral components are following
\begin{eqnarray}\label{E_example_2}
&& {\cal E}_{-5} = -0.1E_{0},\;{\cal E}_{-4} = 0.2E_{0},\;{\cal E}_{-3} = -0.6E_{0},\nonumber\\
&&{\cal E}_{-2} = 0.25E_{0},\;{\cal E}_{-1} = -0.7E_{0},\;{\cal E}_{0} = 0.15E_{0},\\
&&{\cal E}_{+1} = 0.7E_{0},\;{\cal E}_{+2} = 0.5E_{0},\;{\cal E}_{+3} = 0.35E_{0},\nonumber\\
&&{\cal E}_{+4} = 0.1E_{0},\;{\cal E}_{+5}= 0.25E_{0},\nonumber
\end{eqnarray}
where the amplitudes of the resonance frequencies coincide, $|{\cal E}_{-1}|=|{\cal E}_{+1}|= 0.7E_{0}$, i.e. the condition $|\Omega_{-1}|= |\Omega_{+1}|$ is satisfied under equality $d_{eg_1}=d_{eg_2}=d$. In this case, we will consider two different versions of the phase relationships (\ref{phi_1}) and (\ref{phi_2}) from the previous Section~V. The Fig.\,\ref{image:Shift_vs_RR_equal} shows the graphs of the shift of the CPT resonance peak versus the laser field intensity $I\propto |\Omega_{L}|^2$ (where $\Omega_{L}=dE_{0}/\hbar$). As we see, in the case of the exact one-photon resonance $\delta_\text{1-ph}=0$ [see Fig.\,\ref{image:Shift_vs_RR_equal}(a)], the light shift of the CPT resonance almost perfectly coincides with the standard ac Stark shift $\bar{\delta}^{\rm (sh)}_\text{ac}$. However, when the one-photon resonance is violated (i.e. $\delta_\text{1-ph}\neq 0$), a slight difference between the exact shift and the value $\bar{\delta}^{\rm (sh)}_\text{ac}$ can be seen [see Fig.\,\ref{image:Shift_vs_RR_equal}(b)]. This difference is due to the fact that for the field (\ref{E_example_2}) the diagonal matrix elements of the operator $\hat{\widetilde{P}}$ [see (\ref{P_res})] differ from each other, $P_{11}\neq P_{22}$. As a result, there is an imbalance of the pumping channels from the lower states $|g_{1}\rangle$ and $|g_{2}\rangle$ to the upper state $|e\rangle$ due to off-resonant components. This imbalance, in turn, leads to a deformation (asymmetry) of the CPT resonance line shape and is the cause of some additional shift of the CPT resonance peak with respect to the ac Stark shift $\bar{\delta}^{\rm (sh)}_\text{ac}$ for $\delta_\text{1-ph}\neq 0$. To our knowledge, this effect has not been previously discussed in the scientific literature.

The considered example, in combination with the results of the previous  Section~V, gives, a clear explanation of the experimental fact, usually observed when using miniature VCSEL lasers in CPT spectroscopy, when microwave current injected to the laser diode is used to form the modulated light. This consists in the fact that it is very rare to find a laser sample for which it is possible to select such a value of microwave current that the light shift of the CPT resonance becomes equal to zero. Enormous interest in searching for such a regime was caused by the results of Ref.\,\cite{Zhu_2000}, where numerical calculations were carried out for the $\Lambda$-system in the case of pure phase modulation at the frequency $f\approx\Delta_{\rm hfs}/2$, when the spectral components with $n=\pm 1$ are resonant. These calculations showed the absence of the light shift of CPT resonance at the phase modulation index $M\approx 2.42$. However, despite numerous efforts in this direction, this idea has not yet been widely implemented in practice using VCSELs. Based on the results of our analysis, we can give the following general explanation for this negative fact. First of all, we found a pronounced amplitude-phase specificity of the spectral composition for pure phase modulation at the frequency $f\approx\Delta_{\rm hfs}/2$, which very well manifests itself within our theoretical approach. This specificity consists in the symmetry of the light spectrum ($|{\cal E}_{-n}|=|{\cal E}_{n}|$) and at the same time the non-diagonal elements of the shift operator $\hat{\widetilde{S}}_{\rm sh}$ are practically absent, $S^{}_{12}=S^{*}_{21}\approx 0$ (the exact equality $S^{}_{12}=S^{*}_{21}= 0$ is a rigorous analytical result in the case of $\delta_\text{1-ph} = 0$). Therefore, the vanishing of the CPT resonance shift at $M=2.42$ corresponds to the equality $\bar{\delta}^{\rm (sh)}_\text{ac}=0$.
However, when using VCSELs, the real picture of the modulated field usually differs very much from the case of pure phase modulation and contains a significant contribution from the amplitude modulation. This leads to a significant asymmetry in the amplitudes of the spectral components (in particular, $|{\cal E}_{-1}|\neq |{\cal E}_{+1}|$), while the phase relationship between different spectral components becomes very unpredictable. As a result, a significant influence of the non-diagonal elements ($S^{}_{12}=S^{*}_{21}\neq 0$) of the operator $\hat{\widetilde{S}}_{\rm sh}$ on the CPT resonance shift is practically guaranteed, with all the ensuing consequences. Thus, finding the VCSEL diode with regime of the light shift suppression for CPT resonance at the modulation frequency $f\approx\Delta_{\rm hfs}/2$ can be considered as a lucky find. Because of such unpredictability of the spectral picture for different VCSEL samples, the practical implementation of the idea \cite{Zhu_2000} in a commercial production of CPT atomic clocks seems very problematic.

However, in the case of using an electro-optical modulator, the spectrum of the output laser field is close enough to the case of pure phase modulation, which allows the implementation of the light shift suppression regime \cite{Zhu_2000} in atomic CPT clocks (e.g. see Ref.\,\cite{Radnatarov_JETPLett_2023}).

\section{Further development of our theoretical model}

In this paper, we numerically simulated the case of a steady-state regime, when the condition $d/dt\,\hat{\tilde{\rho}}=0$ was used in Eq.\,(\ref{rho_res_eq}) and the light shift of the CPT resonance peak was analyzed. However, in real atomic clocks, various methods of forming an error signal are applied to stabilize the modulation frequency $f$ at zero of the error signal. The commonly used methods include: the frequency-jumps method, the method of additional frequency harmonic modulation at a low frequency ($\sim$1-10~kHz) with subsequent Fourier processing of the output signal (so-called phase-locking technique), and the recently proposed phase-jumps method \cite{Basalaev_PRAppl_2020}. Moreover, in addition to the regime of continuous interaction of atoms with laser field (continuous wave spectroscopy), the pulsed interaction regime (Ramsey spectroscopy) is often implemented in practice. In all these cases, the calculations already require the use of dynamic solutions of Eq.\,(\ref{rho_res_eq}) (see \cite{Yudin_PRA_2016,Yudin_OptExp_2017}).

Note that we have developed in detail and numerically verified the case of a simple $\Lambda$-system with non-degenerate energy levels. However, our approach can be extended to the case of real hyperfine atomic levels taking into account magnetic (Zeeman) sublevels and arbitrary polarization (in the general case, elliptical) of laser field. In this case, the general type of equations for the density matrix will remain the same [see equations (\ref{DensityMatrix_red}) and (\ref{rho_res_eq})], while the matrices $\hat{S}_{\rm sh}$ and $\hat{P}$ themselves will already take into account the Zeeman substructure of levels and the corresponding low-frequency coherence. The formal difference from (\ref{DensityMatrix_red}) and (\ref{rho_res_eq}) will be the presence of an additional operator term describing the interaction with the magnetic field, which is usually used to destroy the degeneracy over magnetic sublevels. In contrast to the simple $\Lambda$-system, there will be a family of different CPT resonances, in which the clock CPT resonance between Zeeman sublevels with magnetic quantum number $m=0$ will be in a numerous environment of magnetically sensitive CPT resonances between sublevels with $m\neq 0$. Thus, our approach can also be used in the theory of atomic CPT magnetometers.

We also add that the model we have presented, which takes into account the interaction with two resonant frequency components by exact way, has a quite large margin of safety. In many cases, an even more radical simplification is possible, when absolutely all frequency components are taken into account in the same way within the second-order perturbation theory in the field. Then, instead of Eq.\,(\ref{rho_res_eq}), another equation can be used:
 \begin{align}\label{rho_res_mod}
&\frac{\partial}{\partial t}\hat{\tilde{\rho}} + \hat{\Gamma}\{\hat{\tilde{\rho}}\} = - i[\hat{H}^{(0)}_{\rm res},\hat{\tilde{\rho}}] -i[\hat{\widetilde{S}}_{\rm sh},\hat{\Pi}_g\hat{\tilde{\rho}}\hat{\Pi}_g]-\\
&\qquad\qquad\qquad \{\hat{\widetilde{P}},\hat{\Pi}_g\hat{\tilde{\rho}}\hat{\Pi}_g\}+\hat{\Pi}_e{\rm Tr}\{\hat{\widetilde{P}},\hat{\Pi}_g\hat{\tilde{\rho}}\hat{\Pi}_g\},\nonumber\\
&  {\rm Tr}\{\hat{\tilde{\rho}}\} = \rho_{g_{1}g_{1}} + \rho_{g_{2}g_{2}} + \rho_{ee} = 1,\nonumber
\end{align}
which does not contain the interaction operator $\hat{\widetilde{V}}_{\rm res}$ at all, and the solution has the form
\begin{equation}\label{rho_matrix_mod}
    \hat{\tilde{\rho}} =
        \begin{pmatrix}
        \rho_{ee} & 0 & 0\\
        0 & \rho_{g_{1}g_{1}} & \tilde{\rho}_{g_{1}g_{2}}\\
        0 &\tilde{ \rho}_{g_{2}g_{1}} & \rho_{g_{2}g_{2}}
        \end{pmatrix},
\end{equation}
where optical coherences are absent, $\tilde{\rho}^{}_{eg_{j}}=\tilde{\rho}^{*}_{g_{j}e}=0$ ($j=1,2$).

In this case, the expressions for the operators $\hat{H}^{(0)}_{\rm res}$, $\hat{\widetilde{S}}_{\rm sh}$ and $\hat{\widetilde{P}}$ in Eq.\,(\ref{rho_res_eq}) are somewhat transformed. For the operator $\hat{H}^{(0)}_{\rm res}$ we have
\begin{equation}\label{H0_res_mod}
\hat{H}^{(0)}_{\rm res}=
\begin{pmatrix}
        0 & 0 & 0\\
       0 & \delta_{\rm R} & 0\\
       0 & 0 & 0
        \end{pmatrix}.
\end{equation}
The matrix elements of the operator $\hat{\widetilde{S}}_{\rm sh}$ are defined as
\begin{eqnarray}\label{S_mod}
&& S_{11}=\sum\limits_{n}\frac{\delta^{(1)}_{n}\big|\Omega_{n}^{(1)}\big|^{2}}{\gamma_{\rm opt}^{2} +\big|\delta^{(1)}_{n}\big|^{2}},\quad S_{22}=\sum\limits_{n}\frac{\delta^{(2)}_{n}\big|\Omega_{n}^{(2)}\big|^{2}}{\gamma_{\rm opt}^{2} +\big|\delta^{(2)}_{n}\big|^{2}},\nonumber\\
&&S_{12}=\sum_{n}\frac{\delta^{(1)}_{n} \Omega^{(1)*}_{n}\Omega^{(2)}_{n+N}e^{i(\phi^{}_n-\phi^{}_{n+N})}}{\gamma_{\rm opt}^{2} + \big|\delta^{(1)}_{n}\big|^{2}},
\end{eqnarray}
and the matrix elements of the operator $\hat{\widetilde{P}}$ have the form
\begin{eqnarray}\label{P_mod}
&&P_{11}=\sum\limits_{n}\frac{\gamma_{\rm opt}\big|\Omega_{n}^{(1)}\big|^{2}}{\gamma_{\rm opt}^{2} +\big|\delta^{(1)}_{n}\big|^{2}},\quad P_{22}=\sum\limits_{n}\frac{\gamma_{\rm opt}\big|\Omega_{n}^{(2)}\big|^{2}}{\gamma_{\rm opt}^{2} +\big|\delta^{(2)}_{n}\big|^{2}},\nonumber\\
&&P_{12}=\sum_{n}\frac{\gamma_{\rm opt}^{2} \Omega^{(1)*}_{n}\Omega^{(2)}_{n+N}e^{i(\phi^{}_n-\phi^{}_{n+N})}}{\gamma_{\rm opt}^{2} + \big|\delta^{(1)}_{n}\big|^{2}}.
\end{eqnarray}
In contrast to the formulas (\ref{S1122}), (\ref{S12}), (\ref{P1122}) and (\ref{P12}), the expressions (\ref{S_mod})-(\ref{P_mod}) do not contain the restrictions for the summation index $n$. In this case, the CPT resonance forming itself is determined by the operator $\hat{\widetilde{P}}$.

Moreover, it is easy to extend our approach to the case of counter-propagating waves and relatively large one-photon detunings for all frequency components of the periodically modulated laser field. The expressions obtained in this way can be used for the theoretical description of atomic CPT interferometers based on two-photon transitions between hyperfine levels (for example, in alkali metal atoms).

\section{Conclusion}
In conclusion, we have developed an effective mathematical model for calculating the CPT resonance in a periodically modulated laser field, when the modulation frequency $f$ varies near the fractional part of the hyperfine splitting frequency in the ground state $\Delta_{\rm hfs}/N$ (where $N=1,2,3,...$). In this model, only two frequency components that are most resonant with optical transitions in the $\Lambda$-system are exactly taken into account, while all other (relatively off-resonant) frequency components are taken into account by the second-order perturbation theory in the field. Within this approach, equations were obtained for the atomic density matrix, in which all off-resonant frequency components led to the appearance of two new operators (in the general case, non-diagonal): the shift operator $\hat{S}_{\rm sh}$ and the relaxation operator $\hat{P}$. The adequacy of the presented effective model was verified by numerical calculations of graphs of various dependencies, in which we did not find visual differences from the exact calculations.

Our model provides a clear physical picture of various features of CPT spectroscopy in a periodically modulated laser field, including effects that have not been discussed in the scientific literature before. In particular, we clearly show that the widespread viewpoint that the CPT resonance shift is determined by the usual ac Stark shifts of the lower levels is fundamentally incorrect, since in general the contribution to the field shift of the CPT resonance due to beats at the frequency $\Delta_{\rm hfs}$ between different off-resonant frequency components can be comparable to (or even dominate) with respect to the value of the standard ac Stark shift. Therefore, even if we have detailed information on the amplitudes of the spectral components of the modulated radiation (e.g., using a spectrum analyzer), this is, in general, absolutely insufficient to determine the light shift of the CPT resonance if there is no information on the phase relationship between the different field components. In addition, our model provides an explanation for an experimental fact commonly observed when using VCSEL semiconductor lasers in CPT spectroscopy, where a microwave current at a frequency of $f\approx\Delta_{\rm hfs}/2$ injected to the laser diode is used to generate modulated light. This fact is that it is very rare to find a VCSEL sample for which such a value of the microwave current can be chosen that the light shift of the CPT resonance becomes zero, as was theoretically predicted in \cite{Zhu_2000} for the case of pure phase modulation of the laser field.

The obtained results are important for atomic CPT clocks and magnetometers, as well as atomic interferometers based on two-photon transitions in alkali metal atoms. We see further development of our approach in the development of a similar mathematical model that takes into account the Zeeman structure of hyperfine levels and arbitrary elliptical polarization of a periodically modulated laser field.

This work was supported by the Russian Science Foundation (grant 22-72-10096).

\appendix

\section{Derivation of expressions for operators $\hat{S}_{\rm sh}$ and $\hat{P}$, conditioned by the quadratic contribution of the off-resonant frequency components}

We start with the system of equations for the matrix elements of the density matrix (\ref{EqDensMatrE}). We select in the coherences of the ground state $\rho_{g_{1}g_{2}}$ and $\rho_{g_{2}g_{1}}$ oscillations at the resonant radio frequency $(\omega_{n_{2}} - \omega_{n_{1}})=Nf$:
\begin{equation}\label{rho-coh_osc}
    \begin{aligned}
        &\rho_{g_{1}g_{2}} = e^{-i (\omega_{n_{2}} - \omega_{n_{1}}) t} \tilde{\rho}_{g_{1}g_{2}}= e^{-iNf t}\tilde{\rho}_{g_{1}g_{2}}, \\
        &\rho_{g_{2}g_{1}} = e^{i(\omega_{n_{2}} - \omega_{n_{1}}) t}\tilde{\rho}_{g_{2}g_{1}} = e^{iNf t}\tilde{\rho}_{g_{2}g_{1}}.
    \end{aligned}
\end{equation}
Next, we represent the optical coherences as a series:
\begin{eqnarray}\label{OptCohSum}
    &&\rho_{eg_{1}} = \sum\limits_{n}\rho_{eg_{1}}^{(n)}e^{-i(\omega_{n}t+\phi_{n})},\; \rho_{g_{1}e} = \sum\limits_{n}\rho_{g_{1}e}^{(n)}\,e^{i(\omega_{n}t+\phi_{n})},\nonumber\\
    &&\rho_{eg_{2}} = \sum\limits_{n}\rho_{eg_{2}}^{(n)}e^{-i(\omega_{n}t+\phi_{n})},\,\rho_{g_{2}e} = \sum\limits_{n}\rho_{g_{2}e}^{(n)}\,e^{i(\omega_{n}t+\phi_{n})},\nonumber\\
    &&\rho_{g_{1}e}^{(n)} =\rho_{eg_{1}}^{(n)*}, \quad  \rho_{g_{2}e}^{(n)} =\rho_{eg_{2}}^{(n)*}.
\end{eqnarray}
Substituting the expressions (\ref{rho-coh_osc})-(\ref{OptCohSum}) in Eq.\,(\ref{EqDensMatrE}) and taking into account $n_{2}=n_{1}+N$, we obtain the following system of equations, in which we kept only the terms that do not contain fast oscillations with frequencies $\propto f$
\begin{widetext}
\begin{eqnarray}\label{EqDensMatrEls_F}
    &&\frac{\partial}{\partial t}\rho_{g_{1}g_{1}} + \frac{\Gamma}{2}\left(\rho_{g_{1}g_{1}} -\rho_{g_{2}g_{2}}\right) - \gamma_{1}\rho_{ee} = i \left[\Omega_{n_1}^{(1)\ast}\rho_{eg_{1}}^{(n_{1})} - \rho_{g_{1}e}^{(n_{1})}\Omega_{n_1}^{(1)}\right] +  i\sum\limits_{n\neq n_{1}}\left[\Omega_{n}^{(1)\ast}\rho_{eg_{1}}^{(n)} -\rho_{g_{1}e}^{(n)} \Omega_{n}^{(1)}\right],\nonumber\\
    &&\frac{\partial}{\partial t}\rho_{g_{2}g_{2}} +\frac{\Gamma}{2}\left(\rho_{g_{2}g_{2}} - \rho_{g_{1}g_{1}}\right) - \gamma_{2}\rho_{ee} = i\left[ \Omega_{n_2}^{(2)\ast}\rho_{eg_{2}}^{(n_{2})} -  \rho_{g_{2}e}^{(n_{2})}\Omega_{n_2}^{(2)}\right] +  i\sum\limits_{n \neq n_{2}}\left[\Omega_{n}^{(2)\ast}\rho_{eg_{2}}^{(n)} - \rho_{g_{2}e}^{(n)}\Omega_{n}^{(2)}\right],\\
    &&\frac{\partial}{\partial t}\tilde{\rho}_{g_{1}g_{2}} + (\Gamma - i\delta_{\rm R})\tilde{\rho}_{g_{1}g_{2}} = i\left[\Omega_{n_1}^{(1)\ast}\rho_{eg_{2}}^{(n_{2})} - \rho_{g_{1}e}^{(n_{1})}\Omega_{n_2}^{(2)}\right]e^{i(\phi_{n_1}-\phi_{n_2})}+  i\sum\limits_{n \neq n_{1}}\left[\Omega_{n}^{(1)\ast}\rho_{eg_{2}}^{(n + N)} - \rho_{g_{1}e}^{(n)}\Omega_{n+N}^{(2)}\right]e^{i(\phi^{}_{n}-\phi^{}_{n+N})},\nonumber\\
    &&\frac{\partial}{\partial t}\rho_{eg_{1}}^{(n)} + (\gamma_{\rm opt} - i\delta_{1,n})\rho_{eg_{1}}^{(n)} = i \Omega_{n}^{(1)}(\rho_{g_{1}g_{1}} - \rho_{ee}) + i \Omega_{n+N}^{(2)}e^{i(\phi^{}_{n}-\phi^{}_{n+N})}\tilde{\rho}_{g_{2}g_{1}},\nonumber\\
    &&\frac{\partial}{\partial t}\rho_{eg_{2}}^{(n)} + (\gamma_{\rm opt} - i\delta_{2,n})\rho_{eg_{2}}^{(n)} =  i \Omega_{n}^{(2)}(\rho_{g_{2}g_{2}} - \rho_{ee}) + i \Omega_{n-N}^{(1)}e^{i(\phi^{}_{n}-\phi^{}_{n-N})}\tilde{\rho}_{g_{1}g_{2}},\nonumber\\
    &&\tilde{\rho}_{g_{2}g_{1}} =\tilde{\rho}_{g_{1}g_{2}}^{\ast}, \quad \rho_{g_{1}e}^{(n)} =\rho_{eg_{1}}^{(n)\ast}, \quad  \rho_{g_{2}e}^{(n)} =\rho_{eg_{2}}^{(n)\ast},\quad \rho_{g_{1}g_{1}} + \rho_{g_{2}g_{2}} + \rho_{ee} = 1. \nonumber
\end{eqnarray}
\end{widetext}
The following notations are used in the equations (\ref{EqDensMatrEls_F})
\begin{eqnarray}\label{one-ph_detuning}
    &&\delta^{(1)}_{n} = \omega_{n} - \omega_{eg_{1}},\nonumber\\
    &&\delta^{(2)}_{n} = \omega_{n} - \omega_{eg_{2}},\\
    &&\delta_{\rm R} = Nf - \Delta_{\rm hfs},\nonumber
\end{eqnarray}
where $\delta^{(1)}_{n}$ and $\delta^{(2)}_{n}$ are the one-photon detunings for the frequency components of the field (\ref{E_field}), and $\delta_{\rm R}$ is the two-photon (Raman) detuning.

Let us consider optical coherences associated with off-resonant frequency components of the field. Because of relatively large one-photon detunings for off-resonant frequency components, we can assume with a good accuracy the condition of weak temporal dependence:
\begin{equation}\label{opt-coh-amp_cond}
    |\partial \rho_{eg_{j}}^{(n)}/\partial t| \ll |(\gamma_{\rm opt}+i\delta^{(j)}_{n})\rho_{eg_{j}}^{(n)}|, \quad (n\neq n_{1,2}),
\end{equation}
which allows us to neglect the derivatives $\partial\rho_{eg_{j}}^{(n)}/\partial t$ ($j=1,2$) in the corresponding equations (\ref{EqDensMatrEls_F}). In addition, for laser field, which is used in CPT clocks, there is usually very weak saturation of optical transitions. Therefore, neglecting the population of the excited state $\rho_{ee}$, we find
\begin{eqnarray}\label{Opt-Coh-amp}
    &&\rho_{eg_{1}}^{(n)} \approx i\frac{ \Omega_{n}^{(1)}\rho_{g_{1}g_{1}} +  \Omega_{n+N}^{(2)}e^{i(\phi^{}_{n}-\phi^{}_{n+N})}\tilde{\rho}_{g_{2}g_{1}}}{\gamma_{\rm opt} - i\delta^{(1)}_{n}}, \quad (n \neq n_{1}), \nonumber\\
    &&\rho_{eg_{2}}^{(n)} \approx i\frac{ \Omega_{n}^{(2)}\rho_{g_{2}g_{2}} + \Omega_{n-N}^{(1)}e^{i(\phi^{}_{n}-\phi^{}_{n-N})}\tilde{\rho}_{g_{1}g_{2}}}{\gamma_{\rm opt} - i\delta^{(2)}_{n}}, \quad (n \neq n_{2}), \nonumber\\
    &&\rho_{g_{1}e}^{(n)} =\rho_{eg_{1}}^{(n)\ast}, \quad  \rho_{g_{2}e}^{(n)} =\rho_{eg_{2}}^{(n)\ast}.
\end{eqnarray}
Moreover, because of large one-photon detunings for off-resonant components, one can safely ignore the small deviation of the value $Nf$ from the hyperfine splitting $\Delta_{\rm hfs}$, i.e. in (\ref{Opt-Coh-amp}) we will use the following expressions for the detunings
\begin{equation}\label{1ph-detuning}
    \begin{aligned}
        &\delta^{(1)}_{n}=\omega+\frac{n}{N}\Delta_{\rm hfs}-\omega_{eg_1}, \quad (n \neq n_{1}),\\
        &\delta^{(2)}_{n}=\omega+\frac{n}{N}\Delta_{\rm hfs}-\omega_{eg_2}, \quad (n \neq n_{2}),
    \end{aligned}
\end{equation}
which leads to the relationship
\begin{equation}\label{1ph-detuning_rel}
 \delta^{(2)}_{n + N} =\delta^{(1)}_{n},\quad (n \neq n_{1}).
\end{equation}
Substituting the expressions (\ref{Opt-Coh-amp}) in Eq.\,(\ref{EqDensMatrEls_F}) and taking into account (\ref{1ph-detuning})-(\ref{1ph-detuning_rel}), we obtain the equations
\begin{widetext}
\begin{align}\label{EqDensMatrEls_F-nr}
    &\frac{\partial}{\partial t}\rho_{g_{1}g_{1}} + \frac{\Gamma}{2}\left(\rho_{g_{1}g_{1}} - \rho_{g_{2}g_{2}}\right) - \gamma_{1}\rho_{ee} = i \left[\Omega_{n_1}^{(1)\ast}\rho_{eg_{1}}^{(n_{1})} - \Omega_{n_1}^{(1)}\rho_{g_{1}e}^{(n_{1})}\right] - \left[\sum\limits_{n\neq n_{1}}\frac{2\gamma_{\rm opt}|\Omega_{n}^{(1)}|^2}{\gamma_{\rm opt}^{2} + |\delta_{n}^{(1)}|^2}\right]\rho_{g_{1}g_{1}}- \nonumber\\
        &\quad \left[\sum\limits_{n\neq n_{1}}\frac{(\gamma_{\rm opt} + i\delta_{n}^{(1)})\Omega_{n}^{(1)*}\Omega_{n+N}^{(2)}e^{i(\phi^{}_{n}-\phi^{}_{n+N})}}{\gamma_{\rm opt}^{2} + |\delta_{n}^{(1)}|^2}\right]\tilde{\rho}_{g_{2}g_{1}} -\left[\sum\limits_{n\neq n_{1}}\frac{(\gamma_{\rm opt} - i\delta_{n}^{(1)})\Omega_{n}^{(1)}\Omega_{n+N}^{(2)\ast}e^{-i(\phi^{}_{n}-\phi^{}_{n+N})}}{\gamma_{\rm opt}^{2} + |\delta_{n}^{(1)}|^2}\right]\tilde{\rho}_{g_{1}g_{2}} ,\nonumber
\end{align}
\begin{align}
    &\frac{\partial}{\partial t}\rho_{g_{2}g_{2}} +\frac{\Gamma}{2}\left(\rho_{g_{2}g_{2}} - \rho_{g_{1}g_{1}}\right) - \gamma_{2}\rho_{ee} = i\left[ \Omega_{n_2}^{(2)\ast}\rho_{eg_{2}}^{(n_{2})} - \Omega_{n_2}^{(2)}\rho_{g_{2}e}^{(n_{2})}\right] -
    \left[\sum\limits_{n \neq n_{2}}\frac{2\gamma_{\rm opt}|\Omega_{n}^{(2)}|^{2}}{\gamma_{\rm opt}^{2} + |\delta_{n}^{(2)}|^2}\right]\rho_{g_{2}g_{2}}-\nonumber\\
        &\quad \left[\sum\limits_{n\neq n_{1}} \frac{(\gamma_{\rm opt} + i \delta_{n}^{(1)})\Omega_{n}^{(1)}\Omega_{n+N}^{(2)\ast}e^{-i(\phi^{}_{n}-\phi^{}_{n+N})}}{\gamma_{\rm opt}^{2} + |\delta_{n}^{(1)}|^{2}}\right]\tilde{\rho}_{g_{1}g_{2}} - \left[\sum\limits_{n\neq n_{1}} \frac{(\gamma_{\rm opt} - i \delta_{n}^{(1)})\Omega_{n}^{(1)*}\Omega_{n+N}^{(2)}e^{i(\phi^{}_{n}-\phi^{}_{n+N})}}{\gamma_{\rm opt}^{2} + |\delta_{n}^{(1)}|^{2}}\right]\tilde{\rho}_{g_{2}g_{1}},\nonumber\\
    &\frac{\partial}{\partial t}\tilde{\rho}_{g_{1}g_{2}} + (\Gamma - i\delta_{\rm R})\tilde{\rho}_{g_{1}g_{2}} = i\left[\Omega_{n_1}^{(1)*}\rho_{eg_{2}}^{(n_{2})} - \Omega_{n_2}^{(2)}\rho_{g_{1}e}^{(n_{1})}\right]e^{i(\phi_{n_1}-\phi_{n_2})} -\\
        &\quad  \left[\sum\limits_{n \neq n_{1}}\frac{(\gamma_{\rm opt} + i\delta_{n}^{(1)})|\Omega_{n}^{(1)}|^{2}}{\gamma_{\rm opt}^{2} + |\delta_{n}^{(1)}|^{2}} + \sum\limits_{n \neq n_{2}}\frac{(\gamma_{\rm opt} - i\delta_{n}^{(2)})|\Omega_{2}^{(n)}|^{2}}{\gamma_{\rm opt}^{2} + |\delta_{n}^{(2)}|^{2}}\right]\tilde{\rho}_{g_{1}g_{2}}- \nonumber\\
        &\quad  \left[\sum\limits_{n \neq n_{1}}\frac{(\gamma_{\rm opt} - i\delta_{n}^{(1)})\Omega_{n}^{(1)*}\Omega_{n+N}^{(2)}e^{i(\phi^{}_{n}-\phi^{}_{n+N})}}{\gamma_{\rm opt}^{2} + |\delta_{n}^{(1)}|^{2}}\right]\rho_{g_{1}g_{1}} - \left[\sum\limits_{n \neq n_{1}}\frac{(\gamma_{\rm opt} + i\delta_{n}^{(1)})\Omega_{n}^{(1)*}\Omega_{n+N}^{(2)}e^{i(\phi^{}_{n}-\phi^{}_{n+N})}}{\gamma_{\rm opt}^{2} + |\delta_{n}^{(1)}|^{2}}\right]\rho_{g_{2}g_{2}},\nonumber\\
    &\frac{\partial}{\partial t}\rho_{eg_{1}}^{(n_{1})} + (\gamma_{\rm opt} - i\delta^{(1)}_{n_1})\rho_{eg_{1}}^{(n_{1})} = i \Omega_{n_1}^{(1)}(\rho_{g_{1}g_{1}} - \rho_{ee}) + i \Omega_{n_2}^{(2)}e^{i(\phi_{n_1}-\phi_{n_2})}\tilde{\rho}_{g_{2}g_{1}},\nonumber\\
    &\frac{\partial}{\partial t}\rho_{eg_{2}}^{(n_{2})} + (\gamma_{\rm opt} - i\delta^{(2)}_{n_{2}})\rho_{eg_{2}}^{(n_{2})} = i \Omega_{n_2}^{(2)}(\rho_{g_{2}g_{2}} - \rho_{ee}) + i \Omega_{n_1}^{(1)}e^{-i(\phi_{n_1}-\phi_{n_2})}\tilde{\rho}_{g_{1}g_{2}},\nonumber\\
    &\tilde{\rho}_{g_{2}g_{1}} =\tilde{\rho}_{g_{1}g_{2}}^{\ast}, \quad \rho_{g_{1}e}^{(n_{1})} =\rho_{eg_{1}}^{(n_{1})\ast}, \quad  \rho_{g_{2}e}^{(n_{2})} =\rho_{eg_{2}}^{(n_{2})\ast},\quad \rho_{g_{1}g_{1}} + \rho_{g_{2}g_{2}} + \rho_{ee} = 1. \nonumber
\end{align}
\end{widetext}
Next, we separate the contributions from the off-resonant components in (\ref{EqDensMatrEls_F-nr}) as follows: the fractional terms in the first three equations whose numerator is proportional to $\pm i\delta^{(1)}_{n}$ or $\pm i\delta^{(2)}_{n}$ (for $n\neq n_{1,2}$) will determine the body of the new operator $\hat{S}_{\rm sh}$, and the fractional terms with a numerator proportional to $\gamma_{\rm opt}$ will determine the body of another operator $\hat{P}$. As a result, passing back to the matrix elements $\rho_{g_{1}g_{2}}$ and $\rho_{g_{2}g_{1}}$ [see (\ref{rho-coh_osc})], and to the original notations for optical coherences, formally setting $\rho^{}_{eg_j}=\rho^{*}_{g_je}=e^{-i(\omega_{n_j}t+\phi_{n_j})}\rho_{eg_{j}}^{(n_{j})}$ (for $j=1,2$), the system of equations (\ref{EqDensMatrEls_F-nr}) can be represented as the following operator equation
\begin{widetext}
\begin{equation}\label{DensityMatrixEq-bichr}
\frac{\partial}{\partial t}\hat{\rho} + \hat{\Gamma}\{\hat{\rho}\} = - i[\hat{H}_{0},\hat{\rho}] + i[\hat{V}_{\rm res},\hat{\rho}]-
 i[\hat{S}_{\rm sh},\hat{\Pi}_{g}\hat{\rho}\hat{\Pi}_{g}] - \{\hat{P},\hat{\Pi}_{g}\hat{\rho}\hat{\Pi}_{g}\} +\hat{\Pi}_{e} {\rm Tr} \{\hat{P},\hat{\Pi}_{g}\hat{\rho}\hat{\Pi}_{g}\}\,,
\end{equation}
where
\begin{equation}\label{Vres}
\hat{V}_{\rm res} = \Omega_{n_1}^{(1)}e^{-i(\omega_{n_{1}}t+\phi_{n_1})}|e \rangle\langle g_{1}|+ \Omega_{n_2}^{(2)}e^{-i(\omega_{n_{2}}t+\phi_{n_2})}|e \rangle\langle g_{2}| + \text{c.c.},
\end{equation}
\begin{eqnarray}\label{ShiftOp}
    &&\hat{S}_{\rm sh} =
    \left[\sum\limits_{n\neq n_{1}}\frac{\delta^{(1)}_{n} |\Omega_{n}^{(1)}|^{2}}{\gamma_{\rm opt}^{2} + |\delta^{(1)}_{n}|^{2}}\right]|g_{1} \rangle\langle g_{1}|+
     \left[\sum\limits_{n\neq n_{2}}\frac{\delta^{(2)}_{n} |\Omega_{n}^{(2)}|^{2}}{\gamma_{\rm opt}^{2} + |\delta^{(2)}_{n}|^{2}}\right]|g_{2} \rangle\langle g_{2}|+\\
    && \left[\sum\limits_{n\neq n_{1}}\frac{\delta^{(1)}_{n} \Omega_{n}^{(1)\ast}\Omega_{n+N}^{(2)}e^{i(\phi^{}_{n}-\phi^{}_{n+N})}}{\gamma_{\rm opt}^{2} + |\delta^{(1)}_{n}|^{2}}\right]e^{-iNft}|g_{1} \rangle\langle g_{2}|+
     \left[\sum\limits_{n\neq n_{1}}\frac{\delta^{(1)}_{n} \Omega_{n}^{(1)}\Omega_{n+N}^{(2)*}e^{-i(\phi^{}_{n}-\phi^{}_{n+N})}}{\gamma_{\rm opt}^{2} + |\delta^{(1)}_{n}|^{2}}\right]e^{iNft}|g_{2} \rangle\langle g_{1}|,\nonumber
\end{eqnarray}
\begin{eqnarray}\label{PopOp}
&&\hat{P} = \left[\sum\limits_{n\neq n_{1}}\frac{\gamma_{\rm opt} |\Omega_{n}^{(1)}|^{2}}{\gamma_{\rm opt}^{2} + |\delta^{(1)}_{n}|^{2}}\right]|g_{1} \rangle\langle g_{1}|+
\left[\sum\limits_{n\neq n_{2}}\frac{\gamma_{\rm opt} |\Omega_{n}^{(2)}|^{2}}{\gamma_{\rm opt}^{2} + |\delta^{(2)}_{n}|^{2}}\right]|g_{2} \rangle\langle g_{2}|+ \\
&&\left[\sum\limits_{n\neq n_{1}}\frac{\gamma_{\rm opt}\Omega_{n}^{(1)\ast}\Omega_{n+N}^{(2)}e^{i(\phi^{}_{n}-\phi^{}_{n+N})}}{\gamma_{\rm opt}^{2} + |\delta^{(1)}_{n}|^{2}}\right]e^{-iNft}|g_{1} \rangle\langle g_{2}|+
\left[\sum\limits_{n\neq n_{1}}\frac{\gamma_{\rm opt} \Omega_{n}^{(1)}\Omega_{n+N}^{(2)*}e^{-i(\phi^{}_{n}-\phi^{}_{n+N})}}{\gamma_{\rm opt}^{2} + |\delta^{(1)}_{n}|^{2}}\right]e^{iNft}|g_{1} \rangle\langle g_{2}|.\nonumber
\end{eqnarray}
\end{widetext}
The description and interpretation of the expressions (\ref{DensityMatrixEq-bichr})-(\ref{PopOp}) are given in Section~III.

\end{document}